\documentclass[draftcls,onecolumn,journal]{IEEEtran}

\usepackage{cite}

\usepackage[cmex10]{amsmath}

\usepackage{array}

\usepackage[pdftex]{graphicx}

\usepackage{amssymb}

\usepackage{caption}
\usepackage{float}

\usepackage{cite}
\usepackage{enumitem}
\usepackage{subfigure}
\usepackage{tabularx}
\usepackage{tabu}

\usepackage{epstopdf}

\usepackage[normalem]{ulem}

\usepackage{booktabs,multirow}

\input{MyMacros.sty}

\begin{document}

\title{Robust Privacy-Utility Tradeoffs\\under Differential Privacy and Hamming Distortion}

%
\author{Kousha Kalantari~\IEEEmembership{Student Member,~IEEE}, 
Lalitha Sankar~\IEEEmembership{Senior Member,~IEEE}, 
and~Anand~D.~Sarwate~\IEEEmembership{Senior Member,~IEEE}
\thanks{Manuscript received January 22, 2018; }%
\thanks{K. Kalantari and L. Sankar are with the  School of Electrical, Computer, and Energy Engineering, Arizona State University, Tempe, Arizona 85287, USA (e-mail: \texttt{kousha.kalantari@asu.edu}, \texttt{lalithasankar@asu.edu}). Their work was supported in part by the National Science Foundation under grant CIF-1422358.}
\thanks{A.D. Sarwate is with the Department of Electrical and Computer Engineering, Rutgers, The State University of New Jersey, Piscataway, NJ 08854, USA (e-mail: \texttt{anand.sarwate@rutgers.edu}). His work was supported in part by the National Science Foundation under grants CCF-1453432, SaTC-1617849, and DARPA and SSC Pacific under contract No. N66001-15-C-4070.}%
\thanks{A preliminary version of these results appeared in the 52nd Annual Allerton Conference~\cite{SarwateSankar} and the 2016 IEEE International Symposium on Information Theory~\cite{KoushaISIT2016}.}
}

\maketitle

\begin{abstract}
A privacy-utility tradeoff is developed for an arbitrary set of finite-alphabet source distributions. Privacy is quantified using differential privacy (DP), and utility is quantified using expected Hamming distortion maximized over the set of distributions. The family of source distribution sets (source sets) is categorized into three classes, based on different levels of prior knowledge they capture. For source sets whose convex hull includes the uniform distribution, symmetric DP mechanisms are optimal. For source sets whose probability values have a fixed monotonic ordering, asymmetric DP mechanisms are optimal. For all other source sets, general upper and lower bounds on the optimal privacy leakage are developed and a necessary and sufficient condition for tightness are established. Differentially private leakage is an upper bound on mutual information (MI) leakage: the two criteria are compared analytically and numerically to illustrate the effect of adopting a stronger privacy criterion.
\end{abstract}


\IEEEpeerreviewmaketitle

\section{Introduction}

The differential privacy (DP) framework offers strong guarantees on the risk of identifying an individual's presence in a database from public disclosures of functions of that database~\cite{DworkMNS:06sensitivity}. This metric has been applied to a variety of computational tasks where privacy guarantees are required, especially in  theoretical computer science, databases, and machine learning. The monograph of Dwork and Roth~\cite{DworkRoth} gives an in-depth treatment of the fundamentals; a short tutorial for signal processing applications introduces basic concepts~\nocite{SarwateC:13survey}. Under differential privacy, randomizing the computation limits the privacy risk, or \emph{leakage}, due to revealing the result of the computation. This randomization often incurs a significant penalty in terms of the usefulness of the published result: this is known as the privacy-utility tradeoff. Differential privacy is a property of the distribution of the computation's output conditioned on its input, which can be modeled information-theoretically as a noisy channel.
In this paper we seek to understand the distortion properties of channels that guarantee differential privacy: this is the privacy-utility tradeoff for the task of publishing a differentially private approximation of the full dataset with utility quantified via a distortion measure.

The DP framework makes no modeling assumptions on the data distribution and gives distribution-independent privacy guarantees. However, in many applications a data published may know something \textit{a priori} about the data distribution. For example, they may know that some elements of the alphabet have a higher probability, or may know something about the distribution up to the labeling of the alphabet.

There are many instances in which a data holder may be required to publish a version of the underlying data. To capture the data holder's knowledge, we assume they know the true distribution lies in an uncertainty set or \emph{source set} of distributions but do not know the true distribution exactly. This knowledge could come from  previously published population statistics, public data or estimation from the source itself, with the uncertainty set represented by confidence intervals. To match the spirit of DP models, we do not assume a Bayesian prior on the set of distributions.

In order to measure the effect of this uncertainty, we model utility as the maximum Hamming distortion over the entire source set. Many datasets contain categorical data for which Hamming distortion is a natural metric and  previously studied privacy mechanisms using additive noise make less sense~\cite{GengV:15opt}: Hamming distortion captures whether the original data was altered or not.
We show that the optimal mechanism guarantees the minimal leakage by effectively censoring low-probability symbols (which are hardest to protect).

\subsection{Our Contributions}

\begin{figure}[h!]
	\centering
	\includegraphics[width=2.4in]{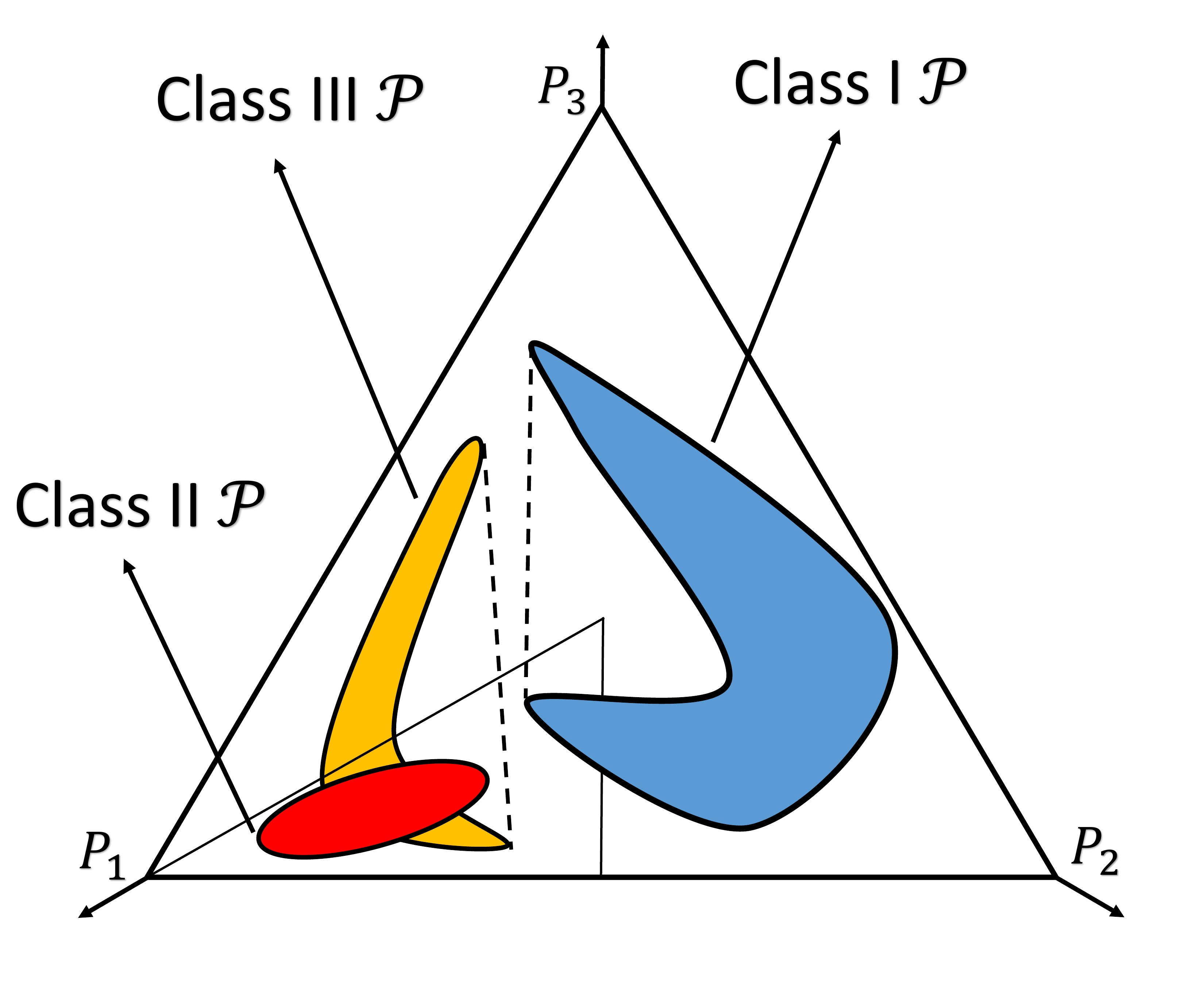}
	\caption{All three classes of source sets for $M=3$.}
	\label{fig: source class types}
\end{figure}

This paper extends our previous results on binary sources~\cite{KoushaISIT2016} to general finite alphabets under Hamming distortion. Larger alphabet sizes permit more complex structures for source sets.
The following are our main contributions:

\begin{enumerate}
\item We categorize source distributions as belonging to one of three possible source classes as illustrated in Figure \ref{fig: source class types} for which different DP mechanisms are optimal.

\begin{itemize}
	\item \textbf{Class I}: Source distribution sets whose convex hull includes the uniform distribution. For example, see the blue source set in Figure \ref{fig: source class types}.
	
	For Class I sources we show that the symmetric mechanism is optimal. Intuitively, this is because knowing that we have a class I source set does not give the data publisher any advantage compared to not knowing anything at all about the source distribution. Therefore, guaranteeing both utility and privacy for all distributions requires a symmetric privacy mapping.
	\item \textbf{Class II}: Source distribution sets that are not Class I, and have ordered probability values. That is, there is a permutation of the alphabet such that all distributions in the class have monotonically decreasing probability mass functions for this permutation. For example, in Figure \ref{fig: source class types}, we have $P_1 \ge P_2 \ge P_3$.
	
We show how to exploit the ordering to characterize optimal non-symmetric mechanisms for the distribution.
As the distortion increases, the optimal mechanism reduces the support size of the output set by mapping low-probability elements to high probability events. We can think of these low-probability events as outliers: since they are most informative (from a privacy perspective), they can be censored in the output to guarantee more privacy.
	
	\item \textbf{Class III}: All source distribution sets that cannot be classified as Class I or Class II. An example is depicted in Figure \ref{fig: source class types}.

We first show that any arbitrary source set can be written as a disjoint union of Class II subsets, each having a different ordering, and use the characterization of Class II mechanisms to derive upper and lower bounds on the privacy leakage.
\end{itemize}

\item We show that the structure of the conditional probability (channel) matrix of optimal mechanisms depends on the location of what we call \textit{critical pairs}: two elements in the same column with maximum ratio.

\item 
We show how the worst-case guarantee of differentially private leakage compares to average-case guarantee of mutual information (MI) leakage. Under standard DP, the mechanism is context-free in the sense that the guarantees do not rely on source distribution assumptions. This leakage upper bounds the context-aware MI leakage, whose guarantees depend on the source distribution~\cite{Sankar_TIFS_2013,CalmonFawaz2013}. This work shows how context-awareness can improve the utility of DP mechanisms and how the gap between the MI leakage and its DP leakage upper bound varies. To do this we study the min-max problems under DP and MI to derive bounds and compute numerical comparisons.
To this end, we formulate the same min-max optimization problem proposed to compute DP mechanisms to determine optimal mechanisms in MI privacy guarantees.
We formulate the same min-max optimization problem using mutual information as the privacy metric. Then, we show that for certain ranges of distortion we can obtain tight bounds and we present numerical comparison.

\end{enumerate}
\subsection{Related Work}

There is a growing body of work on differential privacy (DP) the survey of which is beyond the scope of this paper; we refer the curious reader to the monograph \cite{DworkRoth}. However, comparing differential privacy to other statistical privacy models is a more recent area of study.

Mutual information has been proposed as a metric for privacy leakage~\cite{Rebollo2010,Agrawal2001,Yamamoto} in a variety of settings including data communications, publishing, and mining.
Takbiri et al. \cite{Takbiri-ISIT2017, Takbiri-CISS2017, Takbiri-ISIT17-longversion1} used mutual information to obtain the fundamental limits of privacy in IoT (Internet of Things) devices.
One of the earliest works comparing differential privacy and mutual information privacy is by Alvim et. al.~\cite{Alvim2011,Alvim2012}. Mutual information based privacy metrics have also been considered for data streaming applications~\cite{Khisti2017, Pappas2016, Javidbakht2016}.
Wang et al.~\cite{Wang2016} compare mutual information privacy with differential privacy under Hamming distortion. They also introduce a new privacy measure called identifiability and highlight its relationship to MI and DP.
Building upon prior work~\cite{Alvim2011}, Issa et al.~have introduced maximal leakage (ML) as an information leakage measure for a guessing adversary~\cite{Issa2016}; this measure can also be compared to DP. Cuff and Yu~\cite{Cuff2016} present an equivalent definition of differential privacy using mutual information.

In this paper we consider utility metrics based on rate-distortion in Section \ref{subsection: Information Theoretic Leakage}. A rate-distortion approach to mutual information privacy has been also considered by many researchers~\cite{Sankar_TIFS_2013,CalmonMM15,Asoodeh2015, BWI}; this paper extends our prior work in this direction~\cite{KKSAllerton2016}.


\textit{Local differential privacy} (L-DP)~\cite{Warner,EGS03,KLNRS11,Duchi2013,Guha2017} studies scenarios in which each data respondent, independently applies the same privacy mechanism.
Recently, Kairouz \textit{et al.}~\cite{Kairouz2016} determined the optimal local differential privacy mechanism for a class of utility functions that satisfy a sub-linearity property and show that the resulting L-DP mechanism has a \textit{staircase property}, meaning that the ratio of any two conditional probabilities leading to same outputs is in $\{1,c,c^{-1}\}$, where $c$ is some constant.
However, while Hamming distortion is a sub-linear utility function, the worst case distortion over a source class is not. We show that the staircase property holds only for Class 1 source sets.


\section{Preliminaries and Problem Setup\label{sec:prelims}}

Let $\cX$ represent data value alphabet for each individual. For example, $\cX$ can represent a single attribute or a Cartesian product of other alphabet sets, i.e. $\cX = \prod_{k=1}^{K} \cX_k$, where $\cX_k$ is the set of possible values for $k^{th}$ attribute measured about an individual. Without loss of generality, we assume $\cX = \{1,2,3,\ldots,M\}$. A source distribution set (source set) $\mc{P}$ is a subset of probability simplex on $M$ atoms. We model the data of individuals as being drawn from a distribution $P_X \subset \mc{P}$ on $\cX$, but our solutions can depend on $\mc{P}$ and not the particular $P_X$ is not known \emph{a priori}.

Given an individual's data $X \in \cX$, our goal is to find a conditional distribution $Q_{\hX|X}$ that maps the input data $X$ to an output data $\hX \in \hat{\mathcal{X}}$. Our objective is to find a $Q_{\hX|X}$ that is both \textit{privacy preserving} and does not distort the data above some threshold. For any $P_X \in \mc{P}$, let $Q_{\hX|X}$ and $P_{X,\hX} = P_X Q_{\hX|X}$ indicate the mechanism and the joint distribution of input and output data, respectively. When it is clear from context, we may drop the subscripts and simply use $P$ and $Q$ instead of $P_X$ and $Q_{\hX|X}$. We write $P_i$ for $P_X(X=i)$ for $i \in \cX$.

Let $T$ be a permutation on $\{1, \ldots ,M\}$, such that $T(i)$ is the $i^{\text{th}}$ element in the permuted sequence, and $T^{-1}(j)$ is the index of element $j$ in the permuted sequence. Throughout the sequel, we also refer to the permuted version of a source distribution $P$ as $T(P)$, which is a distribution $\hat{P}$ such that for all $i$, we have $\hat{P}_{T(i)} = P_i$. Likewise, we write $T^{-1}(\mc{P})$ to denote the inverse of $T$ applied to $\mc{P}$.
The notation $T(\mc{P})$ (respectively $T^{-1}(\mc{P})$) is the image of $\mc{P}$ (respectively $T(\mc{P})$) when $T$ (respectively $T^{-1}$) is applied to every $P \in \mc{P}$.

\subsection{Classification of source distribution sets \label{sec:sourceclasses}}

We divide the set of source distribution sets into three classes. 
\begin{definition}
A source set $\mc{P}$ is classified as follows:
\begin{itemize}[leftmargin=2\parindent]
	\item A source set $\mc{P}$ is of \textbf{Class I} if its convex hull $\conv(\mc{P})$ includes the uniform distribution $P_i =\frac{1}{M}$ for all $i \in \{1,2,\ldots,M\}$.
		
	\item A source set $\mc{P}$ is of \textbf{Class II} if it is not Class I and there exists a single permutation $T(\cdot)$ such that for every distribution $P = (P_1, P_2,P_3, \dots, P_M) \in \mc{P}$, we have $P_{T(1)} \ge P_{T(2)} \ge P_{T(3)} \ge \dots \ge P_{T(M)}$.
	
	\item Any other source set is defined to be of \textbf{Class III}.
\end{itemize}
\label{def: classes}
\end{definition}
\begin{remark}
Without loss of generality, for Class II source sets we assume $P_1 \ge P_2 \ge \ldots \ge P_M$.
\label{remark: P_i s are ordered}
\end{remark}

We now provide some examples for source sets in Classes II and III defined above. These examples are used in Section \ref{sec: illustration} to derive numerical comparisons between different source classes.

\begin{example}
In a Class II set, all source distributions have entries (as vectors) ordered in the same way. For example, a $\mc{P}$ containing a single distribution, such as $\mc{P}^{(6)}_{\text{II}}$ shown in Table \ref{table: m 6} for $M = 6$, is a Class II set. A line segment between two distributions with the same order is a Class II set. An example $\mc{P}^{(10)}_{\text{II}}$ for $M = 10$ is given in Table \ref{table: m 10}: the line segment between the two rows is a Class II set.

\begin{table}[h!]
	\centering
	\begin{tabu} to 0.3\textwidth{|| X[c] X[c] X[c] X[c] X[c] X[c] ||} 
		\hline
		$P_1$ & $P_2$ & $P_3$ & $P_4$ & $P_5$ & $P_6$ \\ [0.5ex] 
		\hline\hline
		0.7 & 0.15 & 0.06 & 0.04 & 0.03 & 0.02 \\ 
		\hline
	\end{tabu}
	\caption{$\mc{P}^{(6)}_{\text{II}}$ with $M=6$}
	\label{table: m 6}
\end{table}

\begin{table}[h!]
	\centering
	\begin{tabu} to 0.45\textwidth{||X[c] X[c] X[c] X[c] X[c] X[c] X[c] X[c] X[c] X[c]||} 
		\hline
		$P_1$ & $P_2$ & $P_3$ & $P_4$ & $P_5$ & $P_6$ & $P_7$ & $P_8$ & $P_9$ & $P_{10}$  \\ [0.5ex] 
		\hline\hline	
		0.3 & 0.2 & 0.15 & 0.08 & 0.07 & 0.06 & 0.05 & 0.04 & 0.03 & 0.02\\
		\hline
		0.35 & 0.16 & 0.12 & 0.10 & 0.09 & 0.09 & 0.05 & 0.02 & 0.01 & 0.01\\
		\hline
	\end{tabu}
	\caption{$\mc{P}^{(10)}_{\text{II}}$ with $M=10$}
	\label{table: m 10}
\end{table}
\label{example: class II}
\end{example}

\begin{example}
One way to generate a Class III source set is by taking the unions of Class II sets and (some of) their permutations. Examples based on $\mc{P}^{(6)}_{\text{II}}$ and $\mc{P}^{(10)}_{\text{II}}$ are shown in Tables \ref{table: m 6 2} and \ref{table: m 10 2}.
Specifically, we created three additional distributions for each Class II distribution considered by permuting the entries in three different ways to obtain additional distributions. We label these resulting sets as $\mc{P}^{(i)}_{\text{III: a}}$, $\mc{P}^{(i)}_{\text{III: b}}$, and $\mc{P}^{(i)}_{\text{III: c}}$, where $i \in \{6, 10\}$.

Thus, for example, to generate $\mc{P}^{(6)}_{\text{III: a}}$ we
consider both the original $\mc{P}^{(6)}_{\text{II}}$ and a permuted version of it obtained by swapping the first and second entries of $\mc{P}^{(6)}_{\text{II}}$. 
To obtain $\mc{P}^{(6)}_{\text{III: b}}$, we permute the first and third entries of $\mc{P}^{(6)}_{\text{II}}$ and add this to the set $\mc{P}^{(6)}_{\text{III: a}}$. Finally, the set $\mc{P}^{(6)}_{\text{III: c}}$ is obtained by adding to the set $\mc{P}^{(6)}_{\text{III: b}}$ the new distribution obtained by permuting the first and fourth entries of $\mc{P}^{(6)}_{\text{II}}$.
One can similarly construct the sets $\mc{P}^{(10)}_{\text{III: a}}$,$\mc{P}^{(10)}_{\text{III: b}}$, and $\mc{P}^{(10)}_{\text{III: c}}$ by permuting the first and second, first and third, and first and fourth entries of $\mc{P}^{(10)}_{\text{II}}$, respectively. These sets are highlighted in Tables III and IV with their entries denoted by the groupings a, b, and c. Clearly, for both the $M=6$ and $M=10$ cases, the number of entries increases from (a) to (c) reflecting less and less structured sets.


	\begin{table}[h!]
		\centering
		\begin{tabular}{ p{0.05cm} p{0.05cm} p{0.2cm} | *{6}{c} |}
			\cline{4-9}
			& & & $P_1$ & $P_2$ & $P_3$ & $P_4$ & $P_5$ & $P_6$ \\
			\cline{4-9}
			\multirow{4}{*}{c $\left\{\begin{tabular}{@{\ }l@{}}
				\\  \\  \\ \\
				\end{tabular}\right.$}
			&\multirow{3}{*}{b $\left\{\begin{tabular}{@{\ }l@{}}
				\\  \\  \\
				\end{tabular}\right.$}
			&\multirow{2}{*}{a $\left\{\begin{tabular}{@{\ }l@{}}
				\\  \\ 
				\end{tabular}\right.$} &0.7 & 0.15 & 0.06 & 0.04 & 0.03 & 0.02  \\
			\cline{4-9}
			& & &0.15 & 0.7 & 0.06 & 0.04 & 0.03 & 0.02 \\
			\cline{4-9}
			& & &0.06 & 0.15 & 0.7 & 0.04 & 0.03 & 0.02\\
			\cline{4-9}
			& & &0.04 & 0.15 & 0.06 & 0.7 & 0.03 & 0.02  \\
			\cline{4-9}
		\end{tabular}
		\caption{$\mc{P}^{(6)}_{\text{III: a}}$, $\mc{P}^{(6)}_{\text{III: b}}$, and $\mc{P}^{(6)}_{\text{III: c}}$ with $M=6$.}
		\label{table: m 6 2}
	\end{table}

	\begin{table}[h!]
		\centering
		\begin{tabular}{  p{0.05cm} p{0.05cm} p{0.2cm} | *{10}{p{0.3cm}} |}
			\cline{4-13}
			& & & $P_1$ & $P_2$ & $P_3$ & $P_4$ & $P_5$ & $P_6$ & $P_7$ & $P_8$ & $P_9$ & $P_{10}$ \\
			\cline{4-13}
			\multirow{8}{*}{c $\left\{\begin{tabular}{@{\ }l@{}}
				\\  \\  \\ \\ \\ \\ \\ \\
				\end{tabular}\right.$}
			&\multirow{6}{*}{b $\left\{\begin{tabular}{@{\ }l@{}}
				\\  \\  \\ \\ \\ \\
				\end{tabular}\right.$}
			&\multirow{4}{*}{a $\left\{\begin{tabular}{@{\ }l@{}}
				\\  \\ \\ \\
				\end{tabular}\right.$} &0.3 & 0.2 & 0.15 & 0.08 & 0.07 & 0.06 & 0.05 & 0.04 & 0.03 & 0.02\\
			\cline{4-13}
			& & &0.35 & 0.16 & 0.12 & 0.10 & 0.09 & 0.09 & 0.05 & 0.02 & 0.01 & 0.01\\
			\cline{4-13}
			& & &0.2 & 0.3 & 0.15 & 0.08 & 0.07 & 0.06 & 0.05 & 0.04 & 0.03 & 0.02\\
			\cline{4-13}
			& & &0.16 & 0.35 & 0.12 & 0.10 & 0.09 & 0.09 & 0.05 & 0.02 & 0.01 & 0.01\\
			\cline{4-13}
			& & &0.15 & 0.2 & 0.3 & 0.08 & 0.07 & 0.06 & 0.05 & 0.04 & 0.03 & 0.02\\
			\cline{4-13}
			& & &0.12 & 0.16 & 0.35 & 0.10 & 0.09 & 0.09 & 0.05 & 0.02 & 0.01 & 0.01\\
			\cline{4-13}
			& & &0.08 & 0.2 & 0.15 & 0.3 & 0.07 & 0.06 & 0.05 & 0.04 & 0.03 & 0.02\\
			\cline{4-13}
			& & &0.10 & 0.16 & 0.12 & 0.35 & 0.09 & 0.09 & 0.05 & 0.02 & 0.01 & 0.01\\
			\cline{4-13}
		\end{tabular}
		\caption{$\mc{P}^{(10)}_{\text{III: a}}$, $\mc{P}^{(10)}_{\text{III: b}}$, and $\mc{P}^{(10)}_{\text{III: c}}$ with $M=10$.}
		\label{table: m 10 2}
	\end{table}
\label{example: class II|}
\end{example}

\subsection{Distortion measure}

We measure the distortion between $X$ and $\hX$ using Hamming distortion, i.e. $d(x,y)=1$ if $x\neq y$, and $d(x,y)=0$ if $x= y$.

Since Hamming distortion imposes a penalty when the published data is different from the original, it suffices to limit our search for optimal mechanisms to those with an output support set at most equal to $M$. We formally prove this in Section \ref{sec:opt}.
Hamming distortion is particularly meaningful for categorical data in which there may be no natural metric: any difference captures a semantic difference.

We show later that the output alphabet size is at most $M$ as well, depending on the distortion, so $\cX = \hat{\cX}$.
The average distortion is then given as $\mathbb{E}_{X,\hX}[d(X,\hX)]$. To indicate the dependence of the average distortion on the source distribution and the mechanism we write $\mathbb{E}_{P_X, Q_{\hX|X}}[d(X,\hX)]$. For Hamming distortion, the average distortion is $\sum_{i=1}^{M} P_i (1-Q(i|i))$. Thus, we can simplify the $Q$ matrix by defining $D_i = 1-Q(i|i)$ for all $i$. Henceforth, it suffices to consider mechanisms $Q(j|i)$ with the following form:
{
\[
[Q_{\hX|X}]_{ij}=
\begin{bmatrix}
    1-D_1 & Q(2|1) &  \dots & Q(M|1) \\
    Q(1|2) & 1-D_2 &  \dots & Q(M|2)\\
    \vdots & \vdots&  \ddots& \vdots \\
     Q(1|M)& Q(2|M) & \dots  &  1-D_M
\end{bmatrix}.
\]
}
The sub-matrix of $Q$ induced by rows from $i_\text{min}$ to $i_\text{max}$ and columns from $j_\text{min}$ to $j_\text{max}$ is written as $Q(j_\text{min} : j_\text{max} | i_\text{min} : i_\text{max})$.
\begin{definition}
A mechanism $Q_{\hX|X}$, or equivalently its corresponding distortion set $\{D_i\}_{i=1}^{M}$, is called \textit{$(\mc{P},D)$-valid} if it satisfies the average distortion constraint for every $P_X \in \mc{P}$. The set of all $(\mc{P},D)$-valid mechanisms is
\begin{align}
&\mc{Q}(\mc{P},D) \triangleq \nonumber \left\{Q_{\hX|X} : \mathbb{E}\left[d(X,\hX)\right] \le D, \; \forall P_X \in \mc{P}\right\}.
\label{eq: (P,D) valid mechanism}
\end{align}
\label{definition: (P,D) valid mechanism}
\end{definition}

\subsection{Local differential privacy}

We use the same model for local differential privacy as Kairouz et al.~\cite{Kairouz2016}. We borrow the formalization by Kasiviswanathan et al.~\cite{KLNRS11}, which was based on the randomized response mechanism of Warner~\cite{Warner} and Evfimievski et al.~\cite{EGS03}. It is stronger than non-local privacy~\cite{Dwork} and implies local $\epsilon$-differential privacy~\cite{Kairouz2016}, but allows for columns of $Q$ to be all-$0$:
\begin{definition}
	A mechanism $Q_{\hX|X}$ is \textit{$\epsilon$-differentially private} ($\epsilon$-DP) if
	\begin{equation}
	Q(\hx|x_1) \le e^{\epsilon} Q(\hx|x_2) \: \text{for all } x_1,x_2 \in \cX, \hx \in \hat{\cX},
	\label{eq: DP definition}
	\end{equation}
	and
	\begin{equation}
	\begin{split}
	\edp(Q_{\hX|X}) \triangleq \min \Big\{ \epsilon: Q(\hx|x_1) \le e^{\epsilon} Q(\hx|x_2) \quad \quad \\
	\text{for all } x_1,x_2 \in \cX, \hx \in \hat{\cX} \Big\}.
	\end{split}
	\end{equation}
	\label{definition: DP definition}
\end{definition}
\begin{remark}
Note that the privacy parameter $\epsilon$ does not depend on the source class $\mc{P}$.
\end{remark}

\begin{remark}
For a finite $\epsilon > 0$, an $\epsilon$-differentially private mechanism is such that every column has either all non-zero or all zero entries, i.e. there cannot be a zero and a non-zero entry in the same column. Thus, any mechanism achieving a finite $\edp(\cdot)$ can have $M-k$ non-zero columns and $k$ all-zero columns for some integer $0\le k \le M-1$. Also note that $\edp(Q_{\hX|X}) \ge 0$ for any $Q_{\hX|X}$.
\label{remark: all or none}
\end{remark}
Note that $D=0$ (perfect utility) implies that $\hat{X}=X$, i.e. the optimal mechanism is an identity matrix $Q$ with $\edp(Q) = \infty$. Thus, we focus only on $D>0$ in the sequel.
\begin{lemma}
$\edp(\cdot)$ is quasi-convex in $Q_{\hX|X}$, where quasi-convexity is defined according to \cite[Section 3.4.1]{Boyd}. Equivalently \cite[Section 3.4.2]{Boyd}, for all $Q_1, Q_2$, and $\lambda \in [0,1]$ we have
	\begin{equation}
	\edp(\lambda Q_1 + (1-\lambda) Q_2) \le \max\{\edp(Q_1), \edp(Q_2)\}.
	\end{equation}
	
\label{lemma: convexity of epsilon}
\end{lemma}

The proof is given in Section \ref{proof:lemma: convexity of epsilon}.

From Definitions \ref{definition: (P,D) valid mechanism} and \ref{definition: DP definition}, the minimal achievable $\epsilon$-DP for a given distribution set under Hamming distortion is defined as follows.
\begin{definition}
For a source distribution set $\mc{P}$, and a distortion $D$, where $0 < D \le 1$, let
\begin{equation}
\edpopt(\mc{P},D) \triangleq \min_{Q_{\hX|X} \in \mc{Q(\mc{P},D)}} \edp(Q_{\hX|X}).
\label{eq: epsilon_star}
\end{equation}
Also denote the set of all $Q \in \mc{Q}(\mc{P},D)$ that achieve \eqref{eq: epsilon_star} by $\Qopt (\mc{P},D)$.
\end{definition}

\subsection{Worst-case distortion is not sub-linear\label{sec:counterexample}}
The worst-case distortion for a mechanism $Q$ is $\max_{P \in \mc{P}} \sum_{i=1}^{M} P_i D_i$. Since $\edpopt (\mc{P},D)$ is decreasing in $D$, instead of minimizing leakage for a limited worst-case distortion, we can minimize worst-case distortion for limited leakage. Hence, one can formulate the optimization problem in \eqref{eq: epsilon_star} as 
\begin{align}
\min_{Q \in \mc{Q}_\epsilon} \max_{P \in \mc{P}} \sum_{i=1}^{M} P_i D_i = \max_{Q \in \mc{Q}_\epsilon} U(Q),
\end{align}
where utility $U(Q) =- \max_{P \in \mc{P}} \sum_{i=1}^{M} P_i D_i$, and $\mc{Q}_{\epsilon}$ is the set of all $\epsilon$-DP mechanisms.

Kairouz et al.~\cite{Kairouz2016} show the optimality of staircase mechanisms for $U$ satisfying $U({\gamma}Q)={\gamma}U(Q)$ and $U(Q_1+Q_2) \le U(Q_1)+U(Q_2)$; they call such $U$ \textit{sub-linear}. We show by example that our utility is not sub-linear. Let $M=2$ and consider two mechanisms $Q_1$ and $Q_2$ with distortions $D^{(1)}=\{1,0\}$ and $D^{(2)}=\{0,1\}$ respectively, as well as a source set $\mc{P} = \left\{P^{(1)}=(1,0), P^{(2)}=(0,1) \right\}$. Then:
\begin{align}
\nonumber
&-\max_{P\in \mc{P}} \sum_{i=1}^{M} P_i \left(D^{(1)}_i + D^{(2)}_i\right)\\
& \quad > -\max_{P\in \mc{P}} \sum_{i=1}^{M}P_i D^{(1)}_i - \max_{P\in \mc{P}} \sum_{i=1}^{M} P_i D^{(2)}_i.
\end{align}


\section{Main Results \label{sec:opt}}

In the prior work in~\cite{SarwateSankar}, the authors conjecture that the optimal differentially private (DP) mechanism for a discrete source of alphabet size $M$ and distortion level $D$ is
\begin{equation}
{Q_{D}}(j|i) \triangleq \begin{cases}
1-D, & i= j\\
\frac{D}{M-1},& i\neq j\\
\end{cases}.
\label{eq: Q_D}
\end{equation} 
In the following, we show that the achievable scheme in \eqref{eq: Q_D} is tight for Class I source classes; for the Class II source sets, we exactly characterize $\edpopt (\mc{P},D)$ and show that it matches to that of \eqref{eq: Q_D} for well-defined subsets of $D \in (0,1]$, specifically for high and low utility regimes. Finally, we characterize the optimal leakage for Class III source sets of any other form. Note that, $\edpopt(\mc{P},D)=0$ is achievable for $D \ge \frac{M-1}{M}$ by $Q(j|i) = \frac{1}{M}, 1\le i \le M, 1\le j \le M$, for any source set of any class.

\begin{lemma}
	Under Hamming distortion, the minimal leakage of $\mc{P}$ is the same as the minimal leakage of the convex hull of $\mc{P}$.
	\label{lemma: convexity of P}
\end{lemma}

The proof is given in Section \ref{proof:lemma: convexity of P}. Hence, we can always assume $\mc{P}$ is convex.

The following lemma shows that it suffices to limit our search for optimal mechanisms to only those with output support set sizes at most equal to $M$. Therefore, we focus on only such mechanisms throughout the rest of this article
	\begin{lemma}
		For a source set $\mc{P}$, there exists an optimal mechanism with an output alphabet $\mc{\hX}$ satisfying $|\mc{\hX}| \le M$.
		\label{lemma: max output M}
	\end{lemma}
The proof is given in Section \ref{proof:lemma: max output M}.

	

\begin{lemma}
	For a source set $\mc{P}$ of Class II (without loss of generality let $P_{1} \ge P_{2} \ge \dots \ge P_{M}$ for any $P \in \mc{P}$, there exists an optimal mechanism whose corresponding set of $\{D_i\}$ satisfy $D_{1} \le D_{2} \le \dots \le D_{M}$.
	\label{lemma: reverse order of distortion and probability}
\end{lemma}

The proof is given in Section \ref{proof:lemma: reverse order of distortion and probability}.

\begin{theorem}
For any source set $\mc{P}$ of Class I, we have
	\begin{equation}
	\edpopt(\mc{P},D)= \begin{cases}
	\log (M-1) \frac{1-D}{D}, & D \in [0,\frac{M-1}{M}),\\
	0, & D \in [\frac{M-1}{M},1].
	\end{cases}
\end{equation}
\label{theorem:1}
\end{theorem}

For a full proof see Section \ref{proof:theorem:1}. Since the source set includes the uniform point, i.e. the worst distribution, there is no choice other than applying a  symmetric mechanism as if there is no knowledge available. 

We now proceed to Class II source sets, where there is a known order on the probability of each outcome. For such source sets, we use a coloring argument on the entries of $Q$ to prove specific properties that hold for any optimal mechanism. This, in turn, helps us to reduce the dimension of the feasible space and derive the optimal leakage in terms of a minimization over only $M$ diagonal entries of the mechanism. This is formally stated in the next Theorem.

Since utility is a statistical quantity, the statistical knowledge about the source class can be exploited to obtain a better mechanism than the symmetric one.
As the distortion increases, i.e. lower utility is allowed, the size of the output space can decrease. Conversely, for increasing utility, i.e. decreasing distortion, the output space cannot be smaller than a certain size. This leads to a collection of distortion thresholds $D^{(k)}$ at which an additional decrease in output size becomes optimal.

In addition to this observation, we also use the properties of $\epsilon$-DP, and in fact properties of any optimal mechanism, to reduce the dimension of the feasible space from $M^2$ to just $M$ entries.
\begin{theorem} \label{thm:classII}
For a Class II source set $\mc{P}$ with ordered statistics:
\begin{enumerate}[label=(\alph*),leftmargin=2 \parindent]
	\item There is no $(\mc{P},D)$-valid mechanism with $k$ or more all-zero columns for $D < D^{(k)}$, where $D^{0} \triangleq 0$ and
	\begin{equation}
	D^{(k)}\triangleq \displaystyle \max_{P \in \mc{P}}  \sum_{i=M-k+1}^{M} P_i, \quad  1 \le k \le M.
	\label{eq: D^k}
	\end{equation}
	
	\item The optimal leakage is
\begin{align}
&\edpopt(\mc{P},D)=\nonumber\\
&\begin{cases}
\log \left((M-1) \frac{1-D}{D}\right), & { \begin{array}{l} 0 <D < D^{(1)} \end{array}},\\
\displaystyle \min_{l\in \{0,1,\ldots k\}} \epsilon^{(l)^{*}}_{\text{DP:II}}(\mc{P},D), & {\begin{array}{l}
D^{(k)} \le D < D^{(k+1)},\vspace{0mm}\\
k \in \{1,\ldots, M-2\} 
\end{array}}\\
0, &  { \begin{array}{l} D^{(M-1)} \le D \le 1 \end{array}},
\end{cases}
\label{eq: LB II}
\end{align}
where $\epsilon^{(l)^{*}}_{\text{DP:II}}(\mc{P},D)$ is the minimum leakage achievable over all $(\mc{P},D)$-valid mechanisms with exactly $l$ columns with all zero elements and $M-l$ columns with positive elements, formally defined as

\begin{align}
&\epsilon^{(l)^{*}}_{\text{DP:II}}(\mc{P},D) \triangleq  \nonumber\vspace{-2mm} \\
&\begin{array}{rcl}
\hspace{-8mm}& \displaystyle \min_{\{D_i\}_{i=1}^{M-l}} & \displaystyle \log (M-1-l) \frac{1-\frac{\sum_{i=2}^{M-l}D_i}{M-1-l}}{D_1}\\
\hspace{-8mm}& \text{subject to} & \begin{cases}
\sum_{i=1}^{M-l}P_i D_i \le D - D ^{(l)},  \forall P \in \mc{P},&\\
\sum_{i=1}^{M-l} D_i \le M-1-l, & \\
D_i \in [0,1], \forall 1\le i \le M-l,&\\
\end{cases}
\end{array}
\label{eq: epsilon star k Class II}
\end{align}
and the subscript DP:II in \eqref{eq: LB II} and \eqref{eq: epsilon star k Class II} denote the Class II source set.

\end{enumerate}
\label{theorem:2}
\end{theorem}

For a detailed proof see Section \ref{proof:theorem:2}. Note that each column of a mechanism $Q$ with finite $\edp(Q)$ has elements that are either all zero, or all positive. The proof hinges on the fact that for a Class II source set, where we have a complete knowledge on the order of the source distribution probabilities, only mechanisms with specific number of all-zero columns can be feasible for a given distortion $D$. This limitation, together with some properties of the $\edp(\cdot)$ function, result in a specific structure imposed on the optimal mechanism. Therefore, the dimension of the variable space that we need to optimize over reduces to only $M$ instead of $M^2$. 

We now consider the Class III source sets. We show that the optimal mechanism for Class III source sets can be obtained from results for Class II source sets. As a first step to presenting the main result for Class III source sets, we introduce the following notation and definitions.

\begin{figure}[htp]
		\centering
		\includegraphics[width=2.3in]{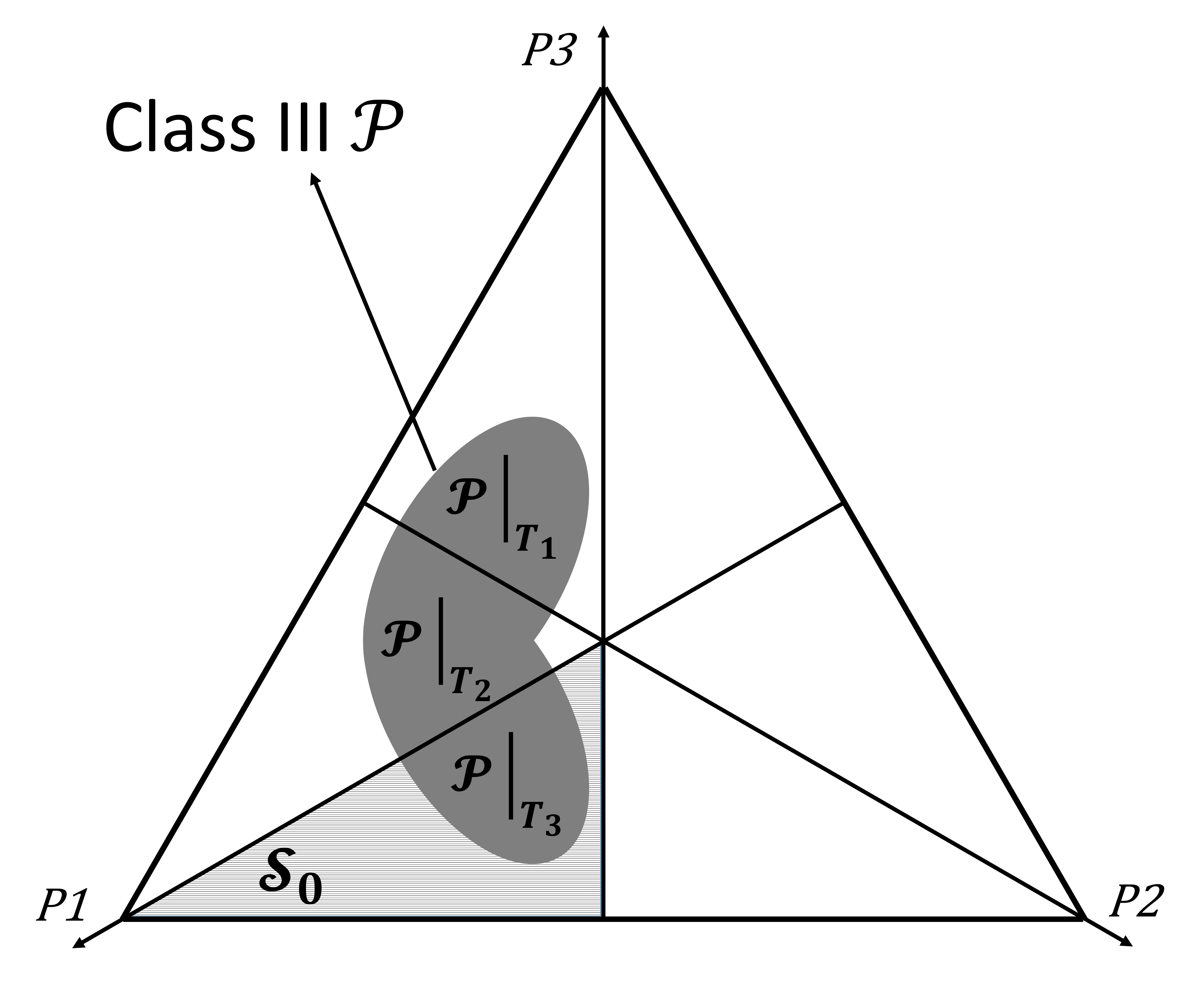}
		\caption{A Class III source set $\mc{P}$.}
		\label{fig: folded 1}
	\end{figure}

\begin{figure}[htp]
		\centering
		\includegraphics[width=0.7\columnwidth]{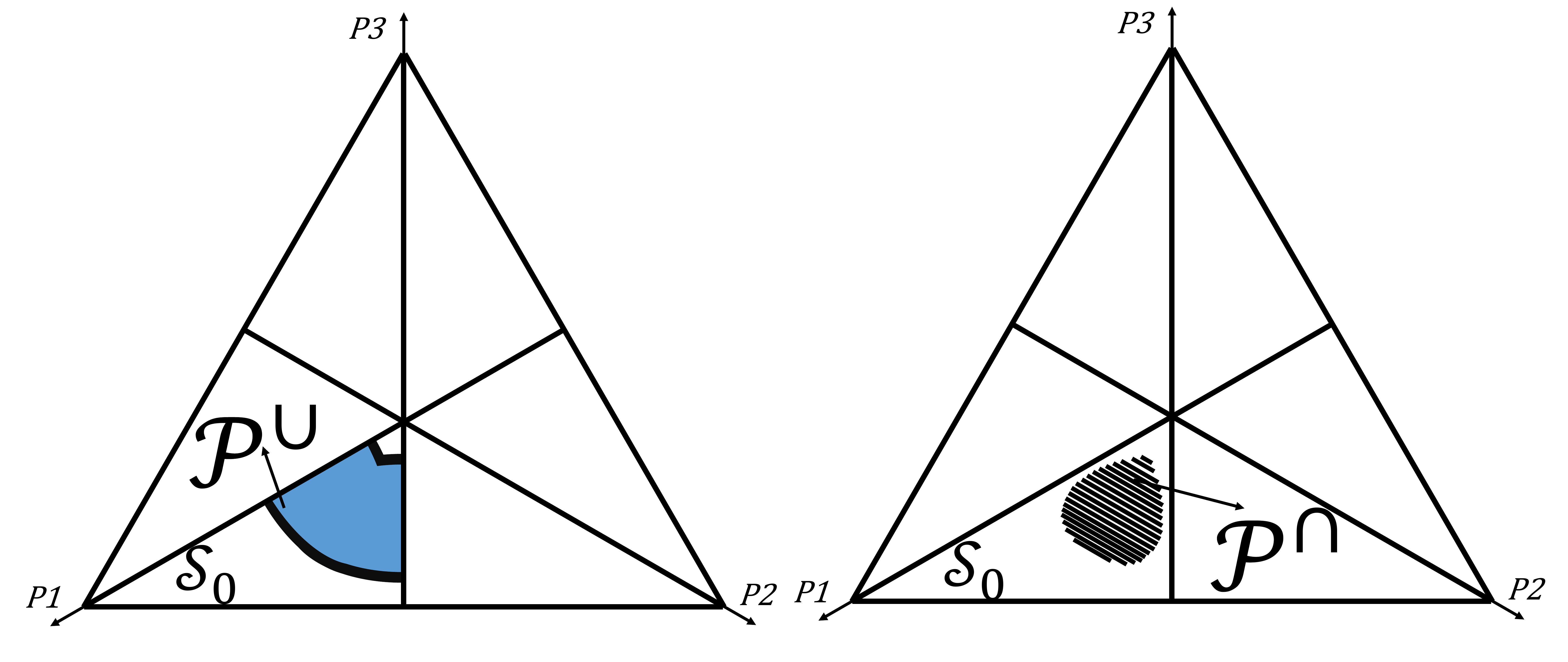}
		\caption{ a Class III source set $\mc{P}$.}
		\label{fig: folded 2}
	\end{figure}

Let $\mc{S}_0$ be the set of all Class II distributions with decreasingly ordered probabilities. Formally,
\begin{equation}
\mc{S}_0 \triangleq \{P: P_1 \ge P_2 \ge \ldots \ge P_M\}.
\end{equation}
Note that the simplex of distributions can be partitioned into $M!$ such ordered subsets, one for each permutation of $\{1,\ldots, M\}$, and thus, there are a total of $M!-1$ other subregions similar to $\mc{S}_0$. For example, for $M=3$, as shown in Figure \ref{fig: folded 1}, the simplex is a union of six disjoint ordered sets. More generally, a subset $\mc{P}$ of the simplex is a union of distributions $P$ that lie in one or more ordered partitions.
For a source distribution $P$ belonging to any one of these partitions, there exists a corresponding folding permutation $T$ such that $P_{T(1)} \ge P_{T(2)} \ge \ldots \ge P_{T(M)}$, or equivalently $T(P) \in \mc{S}_0$. Specifically, any Class III source set $\mc{P}$ can be written as a disjoint union of Class II source sets using what we call \textit{folding permutations}.
\begin{definition}
	Given a Class III source set $\mc{P}$, its \textit{folding permutation set} $\mc{T}_{\mc{P}}$ is the set of all permutations $T$, for which there exists at least one $P \in \mc{P}$ with $P_{T(1)} \ge P_{T(2)} \ge \dots \ge P_{T(M)}$. Then, for each $T \in \mc{T}_{\mc{P}}$ define
	\begin{equation}
	\mc{P}|_{T} \triangleq \{P \in \mc{P}: P_{T(1)} \ge P_{T(2)} \ge \dots \ge P_{T(M)}\}.
	\end{equation}
\end{definition}
Thus, a Class III source set $\mc{P}$ is a union of Class II source sets, i.e. $\mc{P} =  \underset{T \in\mc{T}_{\mc{P}}}{\cup} \mc{P}|_{T}$.
For example, the source set $\mc{P}$ in Fig. \ref{fig: folded 2} lies in three partitions, with corresponding folding permutations $T_1$, $T_2$, and $T_3$. Thus, $\mc{P}|_{T_i}$ is the intersection of $\mc{P}$ with the partition whose folding permutation is $T_i$, $i=1,2,3$, such that $\mc{P} = \mc{P}|_{T_1} \cup \mc{P}|_{T_2} \cup \mc{P}|_{T_3}$.

Without loss of generality, we only focus on those Class III source sets $\mc{P}$ that have a non-empty intersection with $\mc{S}_0$. This is due to the fact that for any other Class III source set, the optimal mechanism can be found using a similar analysis with appropriate change of indices. We now show that for any such Class III source set $\mc{P}$, the optimal leakage can be bounded using the result in Theorem \ref{theorem:2}. We do so by mapping each $\mc{P}|_{T}$ into $\mc{S}_0$, using its corresponding permutation.
\begin{definition}
	For any permutation function $T \in \mc{T}_{\mc{P}}$, we define a \textit{folded equivalent} of $\mc{P}|_{T}$ as its mapped image to $\mc{S}_0$, defined as
	
	
	\begin{equation}
	\overline{\mc{P}|_{T}} \triangleq \left\{\overline{P} \in \mc{S}_0: \exists P \in \mc{P}|_{T} \text{ s.t. } \overline{P} = T(P) \right\}.
	\end{equation}
Furthermore, let
	\begin{equation}
	\mc{P}^{\cap} \triangleq \underset{T \in \mc{T}_{\mc{P}}}{\cap} \overline{\mc{P}|_{T}}, \quad \mc{P}^{\cup} \triangleq \underset{T \in \mc{T}_{\mc{P}}}{\cup} \overline{\mc{P}|_{T}} \label{eq: cap cup}.
	\end{equation}
	\label{def: folding}
\end{definition}
Clearly, $\mc{P}^{\cap} \subseteq \mc{P}^{\cup} \subseteq \mc{S}_0$, and thus, $\mc{P}^{\cap}$ and $\mc{P}^{\cup}$ are Class II source sets. This is depicted in Figure \ref{fig: folded 2}.
	
We now proceed to our main result for Class III source sets.
\begin{theorem}
Let $\mc{P}$ be a Class III source set, such that $\mc{P}^\cap$ and $\mc{P}^\cup$ are non-empty. Then
\begin{equation}
\epsilon_{\text{DP:III}}^{*}(\mc{P}^{\cap},D,\mc{T}_{\mc{P}}) \le \edpopt(\mc{P},D) \le \epsilon_{\text{DP:III}}^{*}(\mc{P}^{\cup},D,\mc{T}_{\mc{P}}),
\label{eq: theorem 3 ub and lb}
\end{equation}
where for any Class II source set $\mc{P}_{\text{II}}$ and a folding permutation set $\mc{T}$ we have:
\begin{align}
&\epsilon_{\text{DP:III}}^{*}(\mc{P}_{\text{II}},D,\mc{T}) \triangleq \nonumber\\
&\begin{cases}
\log \left((M-1) \frac{1-D}{D}\right), & { \begin{array}{l} 0 <D < D^{(1)} \end{array}},\vspace{2mm}\\
\displaystyle \min_{l\in \{0,1,\ldots M\}} \epsilon^{(l)^{*}}_{\text{DP:III}}(\mc{P}_{\text{II}},D,\mc{T}), & {\begin{array}{l}
	D^{(1)} \le D < \frac{M-1}{M},\vspace{0mm} 
	\end{array}}\vspace{3mm}\\
0, &  { \begin{array}{l} \frac{M-1}{M} \le D \le 1 \end{array}},
\end{cases}
\label{eq: LB III}
\end{align}
with
\begin{align}
&\epsilon^{(k)^{*}}_{\text{DP:III}}(\mc{P}_{\text{II}},D,\mc{T}) \triangleq  \nonumber\vspace{-2mm} \\
&\begin{array}{rcl}
\hspace{-5mm} & \displaystyle \min_{\{D_i\}_{i=1}^{M-k}} & \displaystyle \log (M-1-k) \frac{1-\frac{\sum_{i=2}^{M-k}D_i}{M-1-k}}{D_1}\\
\hspace{-5mm} &\text{subject to} & \begin{cases}
\sum_{i=1}^{M-k}P_i D_i \le D - D ^{(k)},& \hspace{4mm} \forall P \in \mc{P}_{\text{II}},\\
\sum_{i=1}^{M-k} D_i  \le M-1-k, &\\
D_{T(i)} = D_i,& \hspace{-11.5mm} \forall \; T \in \mc{T}, 1 \le i \le M, \\
D_i \in [0,1], &  \forall \; 1 \le i \le M.
\end{cases}
\end{array}
\label{eq: epsilon star k Class III}
\end{align}
\label{theorem:3}
\end{theorem}

See Section \ref{proof:theorem:3} for a detailed proof. For any Class III source set, one can determine $\mc{P}^\cap$ and $\mc{P}^\cup$ located inside $\mc{S}_0$, as shown in Figure \ref{fig: folded 2}. The bound results from focusing on $\mc{P}^\cap$ and $\mc{P}^\cup$, and mapping them back using the inverses of all permutations in $\mc{T}_\mc{P}$. The union of all these mapped sets forms $\mc{P}^{\text{LB}}$ and $\mc{P}^{\text{UB}}$, which is contained in and contains $\mc{P}$, respectively. However, $\mc{P}^{\text{LB}}$ and $\mc{P}^{\text{UB}}$ have this specific property that their corresponding leakage can be calculated from applying Theorem \ref{theorem:2} on $\mc{P}^\cap$ and $\mc{P}^\cup$, with an additional constraint of $D_i = D_{T(i)}$, for all $1\le i \le M$ and $T \in \mc{T}_{\mc{P}}$.

\begin{remark}
In the special case where $\mc{P}^{\cap} = \emptyset$, $\edpopt(\mc{P},D)$ can be simply lower bounded by $\edpopt(\mc{P} \cap \mc{S}_0,D)$.
\end{remark}
\begin{remark}
	Observe that $\epsilon^{(k)^{*}}_{\text{DP:II}}$ in \eqref{eq: epsilon star k Class II} and $\epsilon^{(k)^{*}}_{\text{DP:III}}$ in \eqref{eq: epsilon star k Class III} differ in an additional constraint. This comes from the fact that the image of $\mc{P}^{\text{LB}}$ and $\mc{P}^{\text{UB}}$ in each Class II partition is similar, and therefore, forces some distortion values to be equal.
\end{remark}

\begin{corollary}
	For $\mc{P}^{\cap} = \mc{P}^{\cup}$, we have
	\begin{equation}
	\edpopt(\mc{P}^{\cap},D) = \edpopt(\mc{P},D) = \edpopt(\mc{P}^{\cup},D).
	\end{equation}
\end{corollary}
\begin{remark}
If $\mc{P}^{\cap} = \mc{P}^{\cup}$, then the upper and lower bound match, and the minimal leakage is equal to that of $\mc{P}^{\cap}$ obtained by Theorem \ref{theorem:2}, with the additional constraint $D_i = D_{T(i)}$, for all $1\le i \le M$ and $T \in \mc{T}_{\mc{P}}$.
\end{remark}

Finally, note that the solutions provided in Theorem \ref{theorem:2} and Theorem \ref{theorem:3} are found by solving a linear program for fixed $D_1$. This simplifies the optimization considerably for large $M$: a na\"{i}ve exhaustive search over $\Theta(M^2)$ options is reduced to $\Theta(M)$. These formulations are exploited in Section \ref{sec: illustration} to provide intuition comparing Class I, II, and III source sets.


\subsection{Information theoretic leakage}\label{subsection: Information Theoretic Leakage}

Another metric used for leakage is the mutual information between the original and released data, often referred to as \textit{``mutual information (MI) leakage''}. Unlike DP leakage that provides worst case guarantees, MI leakage provides average case guarantees for all entries of a dataset by taking the statistics of the data into account. Another difference between the two is the fact that for any given mechanism, the MI leakage is not only a function of the mechanism $P_{\hat{X}|X}$, but also it is dependent on the specific data distribution $P_X$. For known source distributions, mutual information leakage is studied in~\cite{Sankar_TIFS_2013, CalmonMM15, KKSAllerton2016}, where both asymptotic and non-asymptotic results are derived. However, for the case wherein the source distribution is not known precisely, but some knowledge of source distribution is available, then the worst-case MI leakage of any mechanism $Q$ is defined in~\cite{SarwateSankar} as:
\begin{equation}
\eit(Q) = \max_{P \in \mc{P}} I(P;Q),
\end{equation}
such that the minimal mutual information leakage is defined as
\begin{equation}
\eitopt (\mc{P},D) = \min_{Q \in \mc{Q}(\mc{P},D)} \eit(Q) = \min_{Q \in \mc{Q}(\mc{P},D)} \max_{P \in \mc{P}} I(P;Q).
\end{equation}
Note that it is not in general straightforward to get analytical closed form results for $\eitopt (\mc{P},D)$ for any $\mc{P}$ and a desired utility function.
However, we can characterize its general behavior and use that to make comparisons with DP. Since any mechanism $Q$ that is $(\mc{P},D)$-valid for two source distributions $P_1$ and $P_2$, is also valid for any convex combination of $P_1$ and $P_2$ as well, any source distribution set $\mc{P}$ can be replaced with its convex hull without loss of generality. As a result, the set of valid mechanisms $\mc{Q}(\mc{P},D)$ is also convex. Also note that both $\mc{P}$ and $\mc{Q}(\mc{P},D)$ are compact, i.e. closed and bounded, and mutual information is convex in conditional distribution and convex in source distribution. Therefore, according to the minimax theorem~\cite{minimax} we can conclude that the minimax inequality holds as equality and we have:
\begin{align}
\nonumber
\eitopt (\mc{P},D) &= \min_{Q \in \mc{Q}(\mc{P},D)} \max_{P \in \mc{P}} I(P;Q)\\ 
&= \max_{P \in \mc{P}} \min_{Q \in \mc{Q}(\mc{P},D)} I(P;Q).
\end{align}

We stress that MI leakage and DP leakage 
reflect two very different privacy sensitivity models; in particular, DP leakage is always an upper bound on MI leakage. Therefore, for a common utility function, and a given source set, it is worthwhile to compare their performance. To this end, we present some analytical results under MI leakage for source classes I and II.

\begin{lemma}
For any Class I source set $\mc{P}$, we have
\begin{equation}
\eitopt (\mc{P},D) = 
\begin{cases}
\log M - H(D) - D \log (M-1), & D < \frac{M-1}{M},\\
0, & D \ge \frac{M-1}{M}.
\end{cases}
\end{equation}
\label{lemma: it and dp zero point Class I}
\end{lemma}
\begin{IEEEproof}
We first show that $\eitopt (\mc{P},D) = 0$, if $D \ge \frac{M-1}{M}$. Consider the mechanism which maps every letter of the input alphabet to the first letter. This mechanism results in a distortion of $\frac{M-1}{M}$ and achieves $\eitopt (\mc{P},D) = 0$, because the resulting output distribution is totally independent of the input.

We now proceed to the case where $D < \frac{M-1}{M}$. Recall that $\mc{P}$ is of Class I and includes the uniform point. Since MI leakage is a concave function of $P \in \mc{P}$, for any given mechanism $Q$ the resulting worst case MI leakage is the one corresponding to the uniform source distribution.

The resulting leakage can be lower bounded as
	\begin{align}
	I(X;\hX) &= H(X)-H(X|\hX) \nonumber\\
	& \ge \log M -H(D) - D \log (M-1) \label{eq: lower bound on MI leakage},
	\end{align}
where \eqref{eq: lower bound on MI leakage} follows from Fano's inequality. The lower bound in \eqref{eq: lower bound on MI leakage} can be achieved by the following mechanism:
\begin{equation}
Q(j|i) = \begin{cases}
1-D, & i=j,\\
\frac{D}{M-1}, & i \neq j.
\end{cases}
\end{equation}

\end{IEEEproof}

\begin{lemma}
	For any Class II source set $\mc{P}$, $\eitopt (\mc{P},D) = 0$ iff $D \ge D^{(M-1)}$.
	\label{lemma: it and dp zero point Class II}
\end{lemma}
\begin{IEEEproof}
Let $D \ge D^{(M-1)}$ and $P^* \in \mc{P}$ be the distribution achieving the maximum in definition of $D^{(M-1)}$ in \eqref{eq: D^k}. Consider a mechanism that maps every input independently to the output element of $P^*$ with the highest probability. One can verify that the resulting distortion is $D^{(M-1)} \le D$. Furthermore, one can also verify that for this mechanism $I(X;\hX)=0$.
	
We now show that no mechanism can achieve $I(X;\hX)=0$, if $D < D^{(M-1)}$. Without loss of generality, let $P_1 \ge P_2 \ge \ldots \ge P_M$. Assume to the contrary that there exists a $\hat{Q}$ achieving $\eitopt (\mc{P},D)=0$ for some $D < D^{(M-1)}$. Since $I(X;\hX)=0$, $P(\hx|x)= p(\hx)$ for all $x$. This in turn result in a distortion at least equal to $\sum_{i=2}^{M} P^*_i$, which is equal to $D^{(M-1)}$, and thus, $\hat{Q}$ cannot be $(\mc{P},D)$-valid.
\end{IEEEproof}


\section{Illustration of Results \label{sec: illustration}}



We now illustrate our result by first giving examples of DP leakage for class I,II, and III sources as well as comparisons between DP and MI leakage.
The central question motivating this work is how partial knowledge of the source distribution can be exploited to improve privacy-utility tradeoffs. We revisit our examples from Section \ref{sec:sourceclasses} to illustrate our theoretical results. 

\begin{figure}[h!]
		\centering
		\includegraphics[width= 0.6\columnwidth]{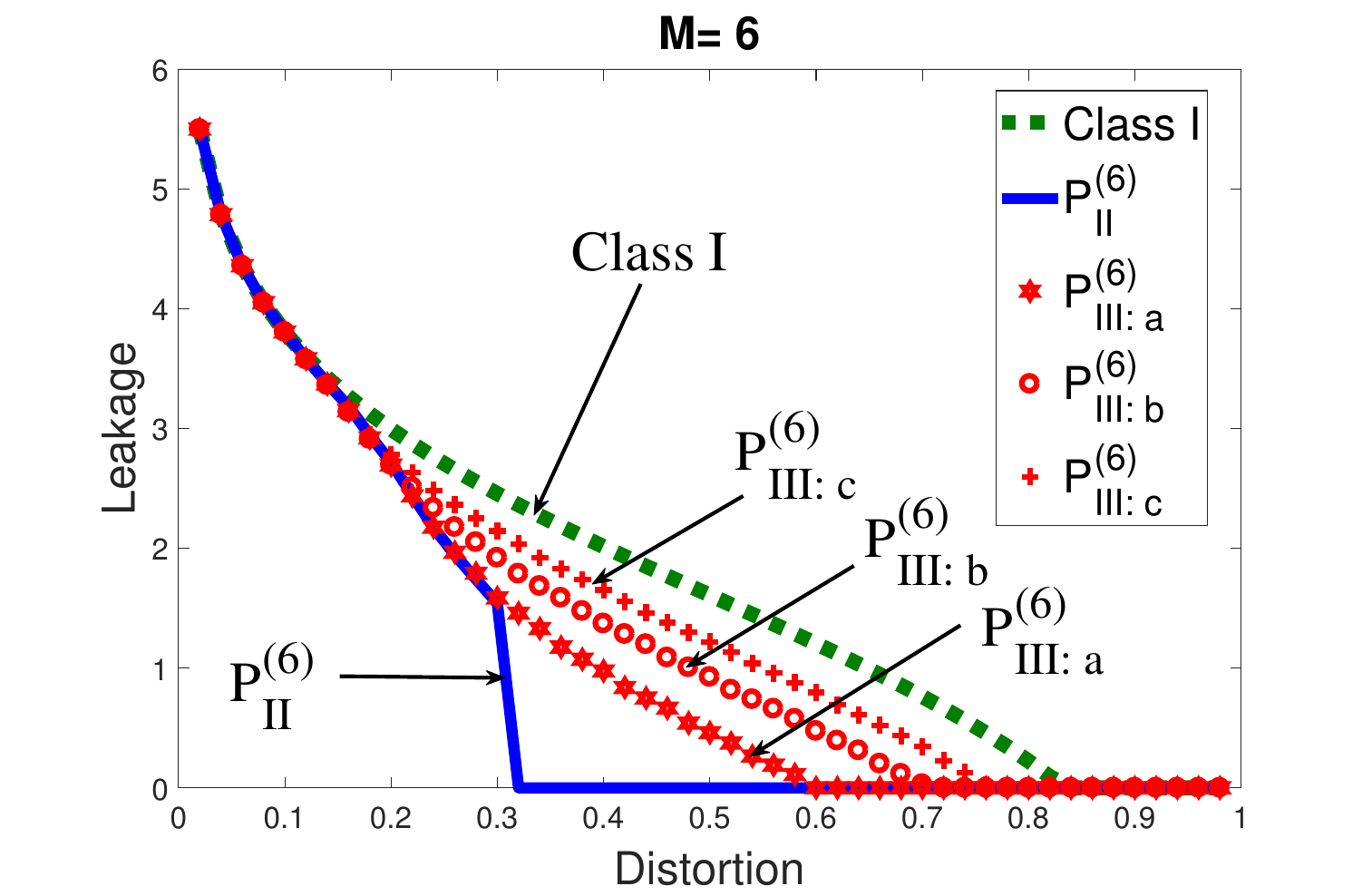}
		\caption{DP leakage-distortion tradeoff for Class I, Class II, and Class III source sets with $M=6$.}
		\label{fig: calssresults 1}
	\end{figure}
\begin{figure}[h!]
		\centering
		\includegraphics[width= 0.6\columnwidth]{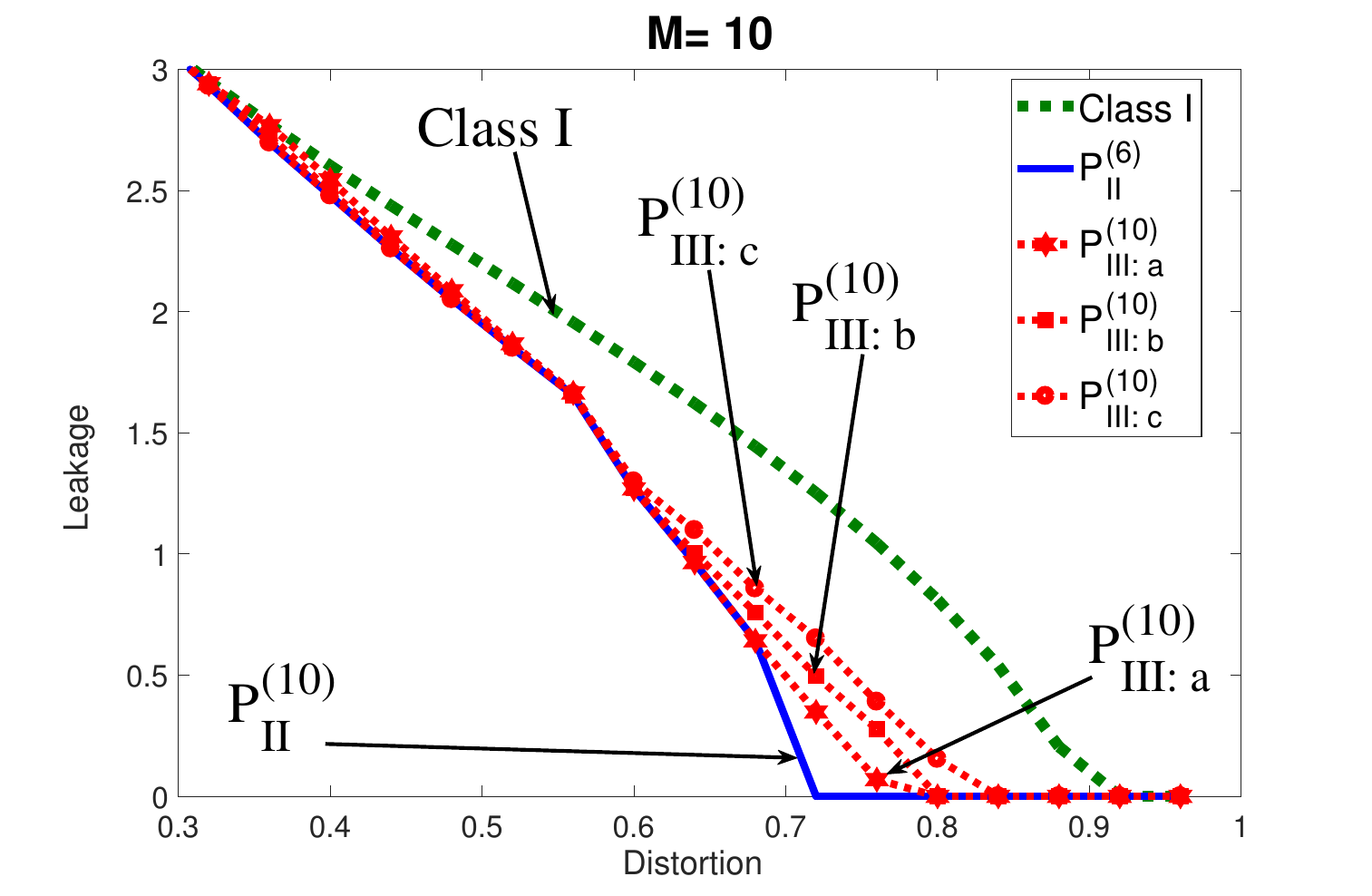}
		\caption{DP leakage-distortion tradeoff for Class I, Class II, and Class III source sets with $M=10$.}
		\label{fig: calssresults 2}
	\end{figure}

We first illustrate the reduction in leakage obtained when one goes beyond Class I source knowledge to Class II. Figures \ref{fig: calssresults 1} and \ref{fig: calssresults 2} show the minimal DP leakage for $\mc{P}^{(6)}_{\text{II}}$ with $M=6$ and $\mc{P}^{(10)}_{\text{II}}$ with $M=10$, respectively. When compared against Class I, for low distortion requirements there is no benefit to source knowledge, but in regimes where a moderate level of distortion is tolerable, the data publisher can significantly decrease the privacy leakage by taking advantage of the source set structure. 

Yet another comparison made here is between Class II and CLass III source sets. Specifically, we expect the leakage guarantees to diminish for Class III which is less structured than Class II and indeed we observe this behavior. 
In fact, since the distortion guarantee for a Class III set also holds for its convex hull, eventually this hull will contain the uniform distribution and the tradeoff will correspond to the Class I leakage. 

\begin{figure}[h!]
		\centering
		\includegraphics[width=0.6 \columnwidth]{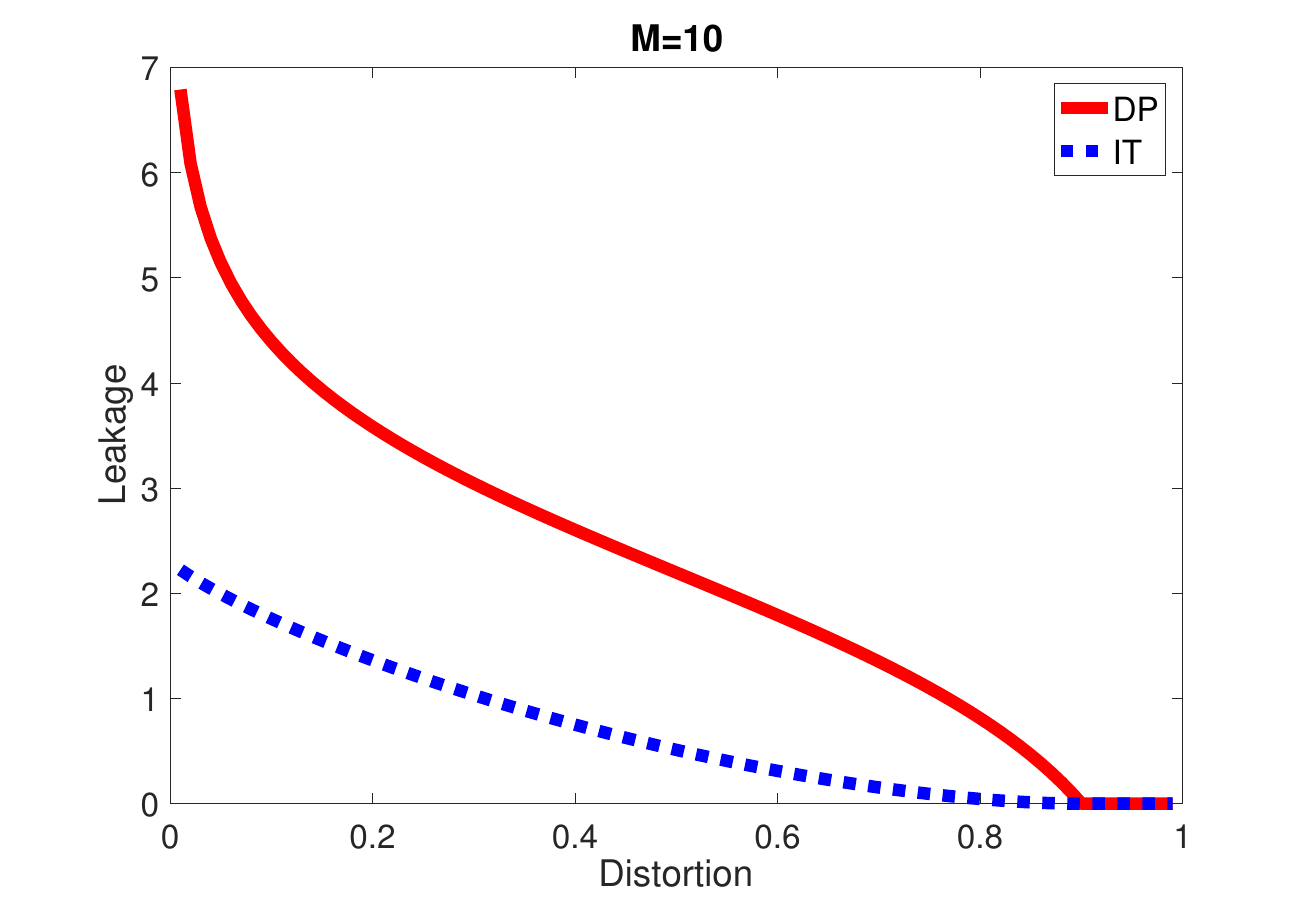}
		\caption{Differential Privacy vs Information Theoretic Leakage for Class I source sets and $M=10$.}
		\label{fig: dp it I}
	\end{figure}

\begin{figure}[h!]
		\centering
		\includegraphics[width=0.6 \columnwidth]{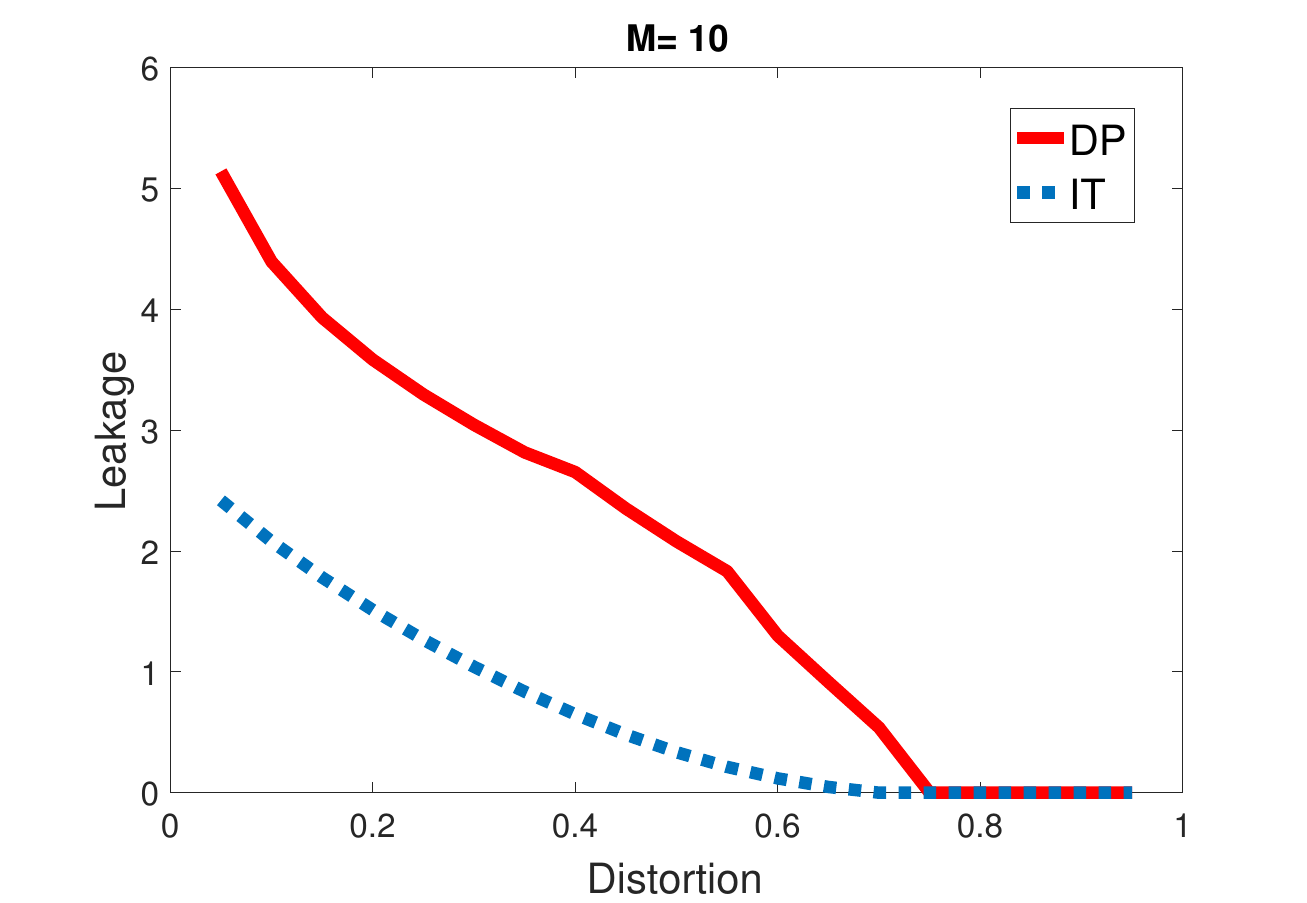}
		\caption{Differential Privacy vs Information Theoretic Leakage for $\mc{P}^{(10)}_{\text{II}}$ in Table \ref{table: m 10}.}
		\label{fig: dp it II}
	\end{figure}

Finally, since DP is distribution-agnostic, the MI leakage is always upper bounded by the DP leakage. In Figures \ref{fig: dp it I} and \ref{fig: dp it II} we compare the MI and DP leakages for Class I and Class II sets. The MI leakage we use is the source-aware worst-case mutual information.

These plots clearly show the convexity of the MI leakage and nonconvexity of the DP leakage as a function of the distortion constraint $D$. The bounds only coincide at perfect privacy, where the output is independent of the input, as indicated by Lemmas \ref{lemma: it and dp zero point Class I} and \ref{lemma: it and dp zero point Class II}.

\section{Proofs}

\subsection{Proof of Lemma \ref{lemma: convexity of epsilon}: quasi-convexity of $\edp(\cdot)$ in $Q$} \label{proof:lemma: convexity of epsilon}

\begin{IEEEproof}
Based on the definition of quasi-convexity in~\cite[Section 3.4]{Boyd}, it suffices to show that all the sub-level sets of the function $\edp(\cdot)$ are convex, i.e. if two different mechanisms $Q_1$ and $Q_2$ are $\epsilon$-differentially private mechanisms for some finite $\epsilon$, then their convex combination $Q_{\theta} = \theta Q_1 + (1-\theta)Q_2, \: 0 < \theta < 1$, is also $\epsilon$-differentially private. Let $x_1$ and $x_2$ be two arbitrary input elements, and let $\hx$ be an arbitrary output element. We have
\begin{subequations}
	\begingroup
	\allowdisplaybreaks
\begin{align}
Q_{\theta}(\hx|x_1) &= \theta Q_1(\hx|x_1) + (1-\theta)Q_2(\hx|x_1)\\
&\le \theta Q_1(\hx|x_2) e^{\epsilon} + (1-\theta)Q_2(\hx|x_2) e^{\epsilon}\\
& = e^{\epsilon} Q_{\theta}(\hx|x_2).
\end{align}
	\endgroup
\end{subequations}
Therefore, $Q_{\theta}$ is also $\epsilon$-differentially private, and thus $\edp(\cdot)$ is a quasi-convex function.
\end{IEEEproof}

\subsection{Proof of Lemma \ref{lemma: max output M}} \label{proof:lemma: max output M}

\begin{IEEEproof}
We first show that for any optimal mechanism $P$ with output support set of size $N$, where $N>M+1$, there exists an optimal mechanism $Q$ with output support set of size $N-1$. It suffices to build $Q$ from $P$ by merging the last two columns of $P$, i.e. adding them element-wise to make one single column. One can verify that $\edp(Q) \le \edp(P)$ due to quasi-convexity of $\edp(\cdot)$ shown in Lemma \ref{lemma: convexity of epsilon}. Note that the resulting distortion is exactly identical in both $Q$ and $P$ since their diagonal elements are equal.
\end{IEEEproof}

We now show that for an optimal mechanism $P$ with output support set of size $M+1$, we can construct an optimal mechanism $Q$ with output support set of size $M$. Take columns $M$ and $M+1$ of $P$, and merge them similar to the previous part. One can similarly verify $\edp(Q) \le \edp(P)$. We now check distortion feasibility. Note that $P_{M,M}=1-D_M$, and therefore once $P_{M,M+1}$ is added to it to obtain $Q_{M,M}$, the updated $D_M$ does not increase, and therefore the total distortion under $Q$ is at most equal to that of $P$. This holds for any distribution point in $\mathcal{P}$.

\subsection{Proof of Lemma \ref{lemma: convexity of P}: Convexity of $\mc{P}$}\label{proof:lemma: convexity of P}

\begin{IEEEproof}
This is due to the fact that any $(\mc{P},D)$-valid mechanism should be also valid for any $\bar{P}$ that is a convex combination of distributions in $\mc{P}$. More formally, suppose that a $\bar{P}$ is in the form of $\sum_{i=1}^{r} \theta_i P^{(i)}$, where $P^{(i)} \in \mc{P}$, for all $i\in \{1,2,\ldots,r\}$. Then, for a $(\mc{P},D)$-valid mechanism $Q$ with distortion set $\{D_i\}$ we have
\begin{subequations}
	\begingroup
	\allowdisplaybreaks
\begin{align}
\sum_{i=1}^{M}\bar{P}_i D_i &= \sum_{i=1}^{M}\sum_{j=1}^{r} \theta_j P^{(j)}_i D_i\\
& = \sum_{j=1}^{r} \theta_j \sum_{i=1}^{M}  P^{(j)}_i D_i\\
& \le \sum_{j=1}^{r} \theta_j D= D.
\end{align}
	\endgroup
\end{subequations}
Thus, $Q$ is also a valid mechanism for $\bar{P}$ and any $\mc{P}$ can be extended to its convex hull without loss of generality.
\end{IEEEproof}

\subsection{Proof of Lemma \ref{lemma: reverse order of distortion and probability} \label{proof:lemma: reverse order of distortion and probability}}
Without loss of generality, let $P_1 \ge P_2 \ge \ldots \ge P_M$ for any $P \in \mc{P}$. We now show that there exists an optimal mechanism with $Q$ with $D_1 \le D_2 \le \ldots \le D_M$. Let $\hat{Q}_{\hX|X}$ be some optimal mechanism such that $\hat{D}_{T(1)} \le \hat{D}_{T(2)} \le \ldots \le \hat{D}_{T(M)}$, for some permutation $T$. Then, let $Q^*_{\hX|X} = \hat{Q}_{T(\hX) | T(X)}$. Clearly, we have $D^*_1 \le D^*_2 \le \cdots \le D^*_M$ and $\edp(\hat{Q}) = \edp(Q^*)$. Finally, $Q^*_{\hX|X}$ is a $(\mc{P},D)$-valid mechanism, because:
\begin{equation}
\sum_{i=1}^{M} P_i D^*_i  \le  \sum_{i=1}^{M} P_i D^*_{T^{-1}(i)} = \sum_{i=1}^{M} P_i \hat{D}_{i} \le D.
\end{equation}

\subsection{Proof of Theorem \ref{theorem:1}: Class I source sets\label{proof:theorem:1}}

We now determine the optimal mechanism for Class I and show that it is indeed the conjectured mechanism in~\cite{SarwateSankar}. From Lemma \ref{lemma: convexity of P}, we know that we can replace $\mc{P}$ with $\conv(\mc{P})$ without loss of generality, and henceforth our results hold for $\conv(\mc{P})$. We begin by assuming to the contrary that there exists a $(\mc{P},D)$-valid mechanism $Q_{\hX|X}$ with lower risk guarantees than conjectured in~\cite{SarwateSankar}, i.e. $\edp(Q) < \log (M-1)\frac{1-D}{D}$ for $0 < D < \frac{M-1}{M}$.
For any $Q$ with $\edp(Q) < \edp(Q_D)$, we require that 
$Q_{\hX|X}(j|i) > e^{\edp(Q_D)} Q_{\hX|X}(j|j)$,
for at least one pair $(i,j), i\neq j$.
Thus, by summing over all columns in $Q_{\hX|X}$ and recalling that $e^{\edp(Q_{D})} = \frac{1-D}{D} (M-1)$, we have

\begin{subequations}
	\begingroup
	\allowdisplaybreaks
\begin{align}
M &= \sum_{i=1}^M \sum_{j=1}^M Q(j|i)\\
\label{eq: lemma 2: nonzero columns}
&> \sum_{j=1}^M \left[ Q(j|j) + \sum_{i\neq j}\frac{Q(i|i)}{e^{\edp(Q_{D})}}\right]\\
&= \sum_{j=1}^{M} \left[  (1-D_j) + \frac{(M-1) - \sum_{i \neq j} D_i}{e^{\edp(Q_{D})}} \right ]\\
&= M - \sum_{j=1}^M D_j + \frac{M(M-1)}{e^{\edp(Q_{D})}} - \frac{M-1}{e^{\edp(Q_{D})}} \sum_{j=1}^M D_j\\
& = \left( \frac{M-1}{e^{\edp(Q_{D})}} +1\right) \left(M - \sum_{j=1}^M D_j\right)\\
&= \left(\frac{1}{1-D}\right) \left(M - \sum_{j=1}^M D_j\right).
\end{align}
	\endgroup
\end{subequations}
Therefore
\begin{align}
\sum_{j=1}^{M} (1-Q(j|j)) = \sum_{j=1}^{M} D_j &> MD.
\end{align}
This, however, contradicts satisfying the distortion constraint for the uniform distribution.

\subsection{Proof of Theorem \ref{theorem:2}: Class II source sets\label{proof:theorem:2}}

We now prove Theorem \ref{theorem:2}, which exactly characterizes $\edpopt(\mc{P},D)$ for the Class II source sets as introduced in Definition \ref{def: classes}.
Recall that we defined distortion levels $D^{(k)}$ in \eqref{eq: D^k} such that for any $k$, $D^{(k)}$ corresponds to the case wherein at most $k-1$ letters of the input are suppressed and the output alphabet size is at least $M-k$, if $D < D^{(k)}$. On the other hand, for $D \ge D^{(k)}$, the output alphabet size may be suppressed by $k$ or more elements. Through the following lemma, we first prove that perfect privacy, i.e. zero leakage, can be achieved if and only if $D \ge D^{(M-1)}$. 

\begin{lemma}
	$\edpopt(\mc{P},D)=0$ if and only if $D \ge D^{(M-1)}$.
	\label{lemma: smallest D for achieving zero privacy}
\end{lemma}

\begin{IEEEproof}	
We first prove the converse and show that $\edpopt(\mc{P},D)=0$ only if $D \ge D^{(M-1)}$.
Let $Q$ be a $(\mc{P},D)$-valid mechanism with $\edp(Q)=0$. This implies all elements of the $i^{\text{th}}$ column have the same value, namely $a_i$, where $0 \le a_i \le 1$ and $\sum_{i=1}^{M} a_i = 1$. Hence, the corresponding distortion values for $Q$ are $D_i = 1-a_i, \forall i$, where $0 \le D_i \le 1$ and $\sum_{i=1}^{M} D_i = M-1$.
Also, recall that for any distribution $P$, we have $P_1 \ge P_2 \ge \ldots \ge P_M$. Therefore, by replacing $D_1$ with zero and $D_i, i>1$ with one, we can further lower bound the distortion as $\sum_{i=1}^{M} P_i D_i \ge \sum_{i=2}^{M} P_i$. Note that $Q$ is a $(\mc{P},D)$-valid mechanism, and therefore $\sum_{i=2}^{M} P_i \le D$. Taking the maximum over all $P \in \mc{P}$ gives $D^{(M-1)}=\max_{P \in \mc{P}}\sum_{i=2}^{M} P_i \le D$.

For proving the achievability, i.e. $\edpopt(\mc{P},D)=0$ for $D \ge D^{(M-1)}$, consider the mechanism with zero elements everywhere except the first column where all entries are $1$, i.e. $Q(i|j)=0$, if $i>1$, and $Q(i|j)=1$, if $i=1$. This mechanism achieves $\edpopt(\mc{P},D)=0$ and the distortion is bounded by
\begin{equation}
\max_{P \in \mc{P}}\sum_{i=1}^{M} P_i D_i  = \max_{P \in \mc{P}}\sum_{i=2}^{M} P_i = D^{(M-1)} \le D.
\end{equation}
\end{IEEEproof}

We now restrict ourselves to $0 \le D \le D^{(M-1)}$, and in the following collection of lemmas we prove structural conditions on the optimal mechanisms for Class II sources. We first describe the need for different distortion levels, and then provide achievability and converse proofs. In particular, as the distortion increases there are specific distortion values at which the support of output is allowed to shrink more. The following lemma captures this observation precisely.

\begin{lemma}\label{lemma: not less than D^k}
For a $k \in \{1,2,\ldots, M\}$ and $D < D^{(k)}$, no $(\mc{P},D)$-valid mechanism can have an output support size of less than or equal to $(M-k)$. 
\end{lemma}

\begin{IEEEproof}
For any $P \in \mc{P}$, any mechanism with $k$ or more all-zero columns  results in an average distortion $\sum_{i=1}^{M} P_i D_i$, which is strictly greater than $\sum_{i=M-k+1}^{M} P_i$ because at least $k$ elements in the set $\{D_i\}_{i=1}^{M}$ are equal to one. Hence, for $D <  D^{(k)}$, no mechanism with $k$ or more all-zero columns can be $(\mc{P},D)$-valid.
\end{IEEEproof}


Recall that without loss of generality, we can assume a given Class II source set has the ordering $P_1 \ge P_2 \ge \ldots \ge P_M$, for any $P \in \mc{P}$. Then, based on Lemma \ref{lemma: reverse order of distortion and probability}, there exists an optimal mechanism with $D_1 \ge D_2 \ge \ldots \ge D_M$.


Using these lemmas, we now present a converse proof by exploiting the definition of differential privacy. We provide a sequence of properties that any optimal mechanism must satisfy. We can therefore obtain a lower bound on the leakage by minimizing parameters of those properties. Then, we present an achievable scheme by providing a mechanism that achieves the minimum value given by the converse.

\subsubsection{Converse for Theorem \ref{theorem:2}}

We now prove a lower bound on $\edpopt(\mc{P},D)$ for $0 < D < D^{(M-1)}$. We first define \textit{critical pairs} in a matrix and then introduce a matrix coloring scheme to prove specific properties of the optimal mechanism. We illustrate this definition and the properties using Figure \ref{fig: coloring}.

\begin{definition}
For a mechanism $Q_{\hX|X}$ with $\edp(Q_{\hX|X}) > 0$, a \textit{critical pair} in $Q_{\hX|X}$ is a pair of elements $\{Q(k|i),Q(k|j)\}$ in a non-zero column, such that $Q_{\hX|X}(k|i) = \exp( \edp(Q_{\hX|X}) ) Q_{\hX|X}(k|j)$.
\label{def: critical pair}
\end{definition}
Note that there exists at least one critical pair, but in general if there are multiple critical pairs in different columns of a matrix $Q$, they may have different values. However, their ratio needs to be equal to $\exp(\edp(Q))$. Furthermore, note that not all columns may have a critical pair. However, the maximal ratio of two elements in any column is at most $\exp(\edp(Q))$. 

We color the entries of non-zero columns of any given matrix $Q$ black, white or red as follows:
\begin{itemize}
	\item An element is colored black if it is the larger element in a critical pair.
	\item An element is colored red if it is the smaller element in a critical pair.
	\item All other elements are colored white.
\end{itemize}
\begin{remark}
Note that if a black element is decreased (or a red is increased), either the $\edp(\cdot)$ of the matrix has to decrease, or that element can no longer be black (red).
\label{remark: no longer black or red}
\end{remark}

Our proof involves manipulating the elements of $Q$ while maintaining it as a valid mechanism: any change in $Q$ that neither increases a black element nor decreases a red element keeps $\edp (Q_{\hX|X})$ at most equal to its previous value.

\begin{definition}
For any $0 \le k \le M-1$, let $\mc{Q}^{(k)}(\mc{P},D)$ be the set of all $(\mc{P},D)$-valid mechanisms with $k$ all-zero columns. Also let $\mc{Q}^{{*}^{(k)}}(\mc{P},D) \subset \mc{Q}^{(k)}(\mc{P},D)$ be the set of mechanisms with the smallest $\edp(\cdot)$ among all mechanisms in $\mc{Q}^{(k)}(\mc{P},D)$.
\end{definition}


Recall from Section \ref{sec:prelims} that $Q(j_\text{min} : j_\text{max} | i_\text{min} : i_\text{max})$ is the sub-matrix of $Q$ induced by rows from $i_\text{min}$ to $i_\text{max}$ and columns from $j_\text{min}$ to $j_\text{max}$. This notation is used extensively throughout the following lemmas and propositions.
\begin{lemma}
For a given $0 < D < D^{(M-1)}$ and $0 \le k \le M-1$, there exists a $Q^{k}_{\hX|X} \in \mc{Q}^{{*}^{(k)}}(\mc{P},D)$ with the distortions $D_1 \le D_2 \le \ldots \le D_M$ such that
\begin{equation}
\edp(Q^{k}_{\hX|X}) =\log (M-1-k) \frac{1-\frac{\sum_{i=2}^{M-k} D_{i}}{M-1-k}}{D_{1}}.
\label{eq: epsilon_prime}
\end{equation}
\label{lemma: epsilon_prime}
\end{lemma}

\begin{IEEEproof}
Denote those mechanisms $\mc{Q}^{{*}^{(k)}}(\mc{P},D)$ that have the smallest $\sum_{i=1}^M D_i$ by $\mc{Q}^{{*}^{(k)}}_{\text{MS}}(\mc{P},D)$, where MS in the subscript stands for ``Minimum Sum''. Through five sequential claims, we now show that there exists a $Q_{\hX|X} \in \mc{Q}^{{*}^{(k)}}_{\text{MS}}(\mc{P},D)$ with a specific color structure as shown in Figure \ref{fig: coloring} and $\edp(Q_{\hX|X})$ is given by \eqref{eq: epsilon_prime}.

\begin{figure}[h!]
\centering
\includegraphics[width=2.4in]{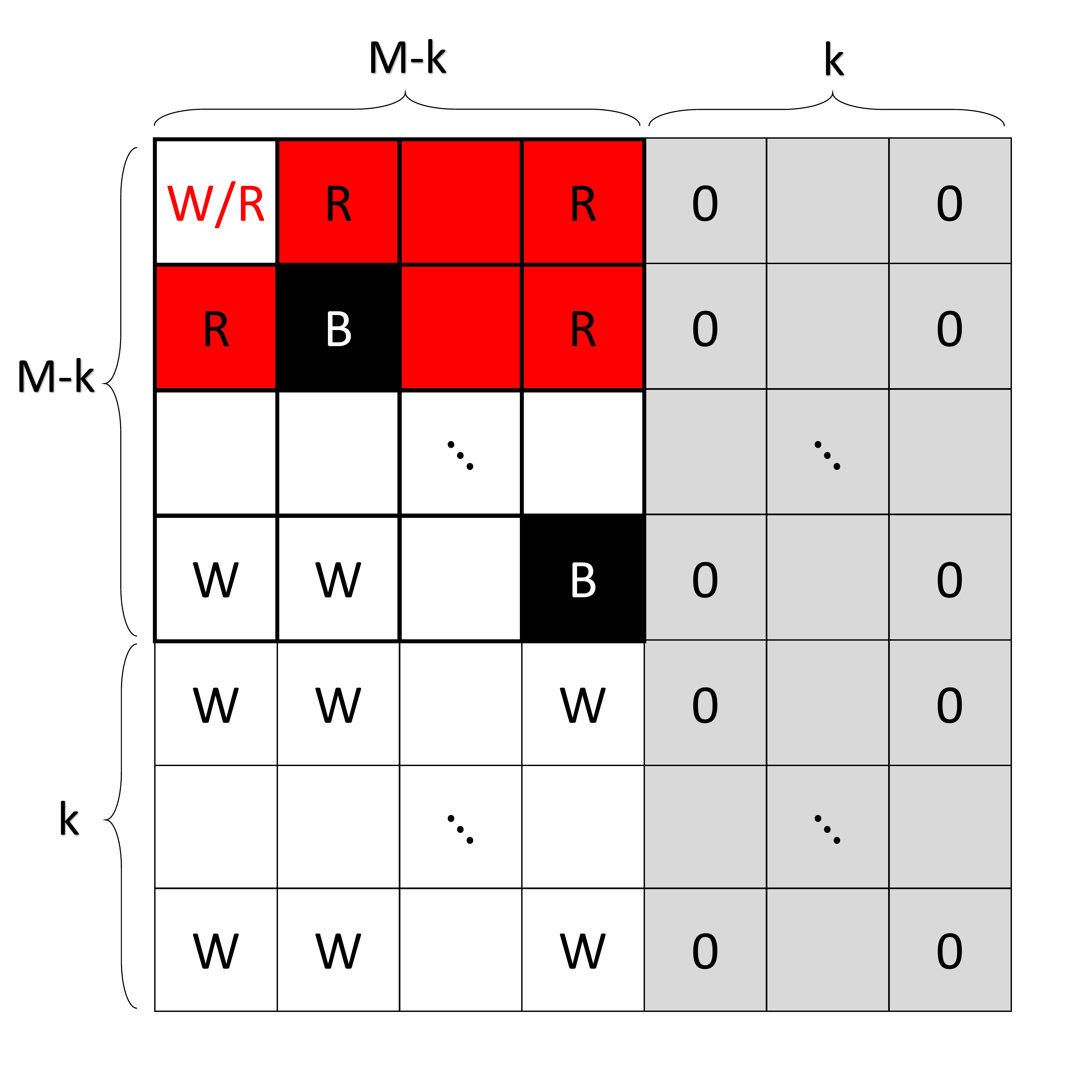}
\caption{For Class II source sets, optimal mechanisms with $k$ all-zero columns have at most one non-black diagonal element, while all other non-zero off diagonal elements are either red or white.}
\label{fig: coloring}
\end{figure}

For any mechanism $Q_{\hX|X}$ with $k$ all-zero columns (or in other words, with $k$ of its $D_i$s being equal to $1$), without loss of generality we can assume that columns $M-k+1$ to $M$ are the all-zero ones due to Lemma \ref{lemma: reverse order of distortion and probability}. Let $Q(1:(M-k) | 1:(M-k))$ and $Q(1:(M-k) | (M-k+1):M)$ be sub-matrices of any given $Q$. Then there exists a $Q_{\hX|X} \in \mc{Q}^{{*}^{(k)}}_{\text{MS}}(\mc{P},D)$ that satisfies the following claims sequentially. In other words, each claim $i$ states that there exists at least one $Q_{\hX|X} \in \mc{Q}^{{*}^{(k)}}_{\text{MS}}(\mc{P},D)$ among all the ones satisfying the previous claims $\{1,\ldots,i-1\}$, such that the statement of claim $i$ is true. 

\begin{claim}
No row is all black (or all red) in $Q(1:(M-k) | 1: M)$. 
\label{claim: 1}
\end{claim}
	
\begin{IEEEproof}
Assume the contrary that the $i^{th}$ row in $Q(1:(M-k) | 1: M)$ is all black for any $Q_{\hX|X} \in \mc{Q}^{{*}^{(k)}}_{\text{MS}}(\mc{P},D)$. Consider the ratio $\edp(Q)=\log\frac{Q(1|i)}{Q(1|j)}$, where $Q(1|j)$ is the red element associated with $Q(1|i)$ which is black, i.e. $Q(1|j)$ and $Q(1|i)$ form critical pairs. Also note that for any other $2 < l \le M-k$, $Q(l|i) \ge Q(l|j)$. Now, if $\edp(Q) > 0$, we have
	\begin{equation}
	1= Q(1|i) +  \sum_{l=2}^{M-k} Q(l|i) > Q(1|j) + \sum_{l=2}^{M-k} Q(l|j) = 1,
	\end{equation}
	which is a contradiction. Otherwise, if $\edp(Q)=0$, by Lemma \ref{lemma: smallest D for achieving zero privacy} $D \ge D^{(M-1)}$, which is also a contradiction to our assumption of $0 < D < D^{(M-1)}$ in the beginning of converse proof. The proof for no row with all red elements is similar.
\end{IEEEproof}

\begin{claim}
All the off-diagonal elements of a row in $Q(1:(M-k) | 1: M)$ have the same color.
	\label{claim: 2}
\end{claim}

\begin{IEEEproof}
	Take an arbitrary mechanism $Q_{\hX|X} \in \mc{Q}^{{*}^{(k)}}_{\text{MS}}(\mc{P},D)$ that $Q(1:(M-k) | 1: M)$ satisfies Claim 1. Fix a row $i$, and let the number of off-diagonal elements in row $i$ of $Q(1:(M-k) | 1: M)$ with each of the colors black, white and red be $n_B$, $n_W$ and $n_R$ respectively, where $n_B + n_W + n_R = M-1-k$. If only one of $n_R$, $n_W$ or $n_B$ is non-zero, then the claim is satisfied. We split the remaining scenarios into two cases and show that in each it is sufficient to make all off-diagonal elements of the $i$th row white.
	
	
\begin{itemize}
		\item If $n_R, n_B >0$, then there exists an arbitrarily small $\delta >0$ such that each of the off-diagonal black elements can be decreased by $\frac{\delta}{n_B}$, and each of the off-diagonal red elements can be increased by $\frac{\delta}{n_R}$.		
Consider a $Q'$ which is identical to $Q$ everywhere except for a segment of the $i$-th row $Q'(1:(M-k)|i)$. For $1\le j \le M-k, j \neq i$, let $Q'(j | i) = Q( j | i ) - \frac{\delta}{n_B}$ if $Q( j | i )$ is black, $Q'(j | i) = Q( j | i ) + \frac{\delta}{n_R}$ if $Q(j | i)$ is red, and $Q'(j|i) = Q(j|i)$ if $Q(j | i)$ is white. Note that it does not matter if $n_W=0$ or not, because the white off-diagonal elements $Q'(1:(M-k)|i)$ are not changed.

		\item If $n_R=0, n_W, n_B >0$ (or $n_B=0, n_W,n_R >0$), then there exists an arbitrarily small $\delta >0$ such that each of off-diagonal white elements can be increased (or decreased) by $\frac{\delta}{n_W}$, and each of off-diagonal black elements (or red elements) can be decreased (or increased) by $\frac{\delta}{n_B}$ (or $\frac{\delta}{n_R}$). Consider a $Q'$ which is identical to $Q$ everywhere except for $Q'(1:(M-k)|i)$. For $1\le j \le M-k, j \neq i$, let $Q'(j | i) = Q( j | i ) - \frac{\delta}{n_B}$ if $Q( j | i )$ is black and $Q'(j | i) = Q( j | i ) + \frac{\delta}{n_W}$ if $Q( j | i )$ is white (or $Q'(j | i) = Q( j | i ) + \frac{\delta}{n_R}$ if $Q( j | i )$ is red and $Q'(j | i) = Q( j | i ) - \frac{\delta}{n_W}$ if $Q( j | i )$ is white), and $Q'(j|i) = Q(j|i)$ elsewhere.
\end{itemize}

In both cases $Q'$ has the same off-diagonal row sum and the same set of $\left\{ D_i\right\}_{i=1}^{M}$ as $Q$, but for sufficiently small $\delta$, $Q'$ is still a valid row stochastic matrix and none of the elements in $Q'(1:(M-k)|1:M)$ become $1$ or $0$. Thus, $Q'$ would still be a $(\mc{P},D)$-valid mechanism. Besides, all off-diagonal elements in row $i$ of $Q'$, $i \notin \{a_1,a_2,\cdots, a_k\}$, are white: If not we would have a smaller $\edp (Q'_{\hX|X})$ due to Remark \ref{remark: no longer black or red}, which contradicts our first assumption that $Q_{\hX|X} \in \mc{Q}^{k^*}(\mc{P},D)$.  In this construction, all off-diagonal elements of $Q'(1:(M-k)|1:M)$ in rows other than $i$ are colored the same as $Q$ without affecting the average distortion, while keeping $\edp(Q') \le \edp(Q)$ and thus $Q' \in \mc{Q}^{{*}^{(k)}}_{\text{MS}}(\mc{P},D)$. This operation can be done for each row $i$ repeatedly, to get the final $Q'$ to satisfy the claim.
	\end{IEEEproof}

\begin{remark}
As a result of Claims \ref{claim: 1} and \ref{claim: 2}, all the off-diagonal elements in $Q(1:(M-k)|(M-k+1):M)$ are white.
\label{remark: lower parts are white}
\end{remark}
		
	\begin{claim}
If a diagonal element in $Q(1:(M-k)|1:(M-k))$ is not black then all off-diagonal elements of $Q(1:(M-k)|1:(M-k))$ in the same row are red.
	\label{claim: 3}
	\end{claim}
	\begin{IEEEproof}
Take an arbitrary $Q_{\hX|X} \in \mc{Q}^{{*}^{(k)}}_{\text{MS}}(\mc{P},D)$ satisfying Claims \ref{claim: 1} and \ref{claim: 2}, and for some $1 \le i \le M-k$ suppose $Q(i|i)$ is red or white. By Claim \ref{claim: 2}, all elements in the set $\{ Q(j | i) : j \ne i, 1\le j \le M-k\}$ have the same color. Assume to the contrary that they are not all red, so they are all black or white. Consider a $Q'$ which is equal to $Q$, except in $Q'(1:(M-k)|i)$ where $Q'(i | i) = Q(i|i) + \delta$ and $Q'(j | i) = Q(j|i) - \frac{\delta}{M-k-1}$ for $j \ne i, 1 \le j \le M-k$. For sufficiently small $\delta > 0$ this is also a $(\mc{P},D)$-valid mechanism. Although $\edp (Q'_{\hX|X})$ remains unchanged due to Remark \ref{remark: no longer black or red}, we have
\begin{equation}
\sum_{i=1}^M D'_i = \sum_{i=1}^{M} (1-Q'(i|i)) < \sum_{i=1}^{M} (1-Q(i|i)) = \sum_{i=1}^M D_i,
\end{equation}
where $D'_i$ is the distortion of $i^{\text{th}}$ element under $Q'$.
This clearly contradicts the assumption that $Q\in \mc{Q}^{{*}^{(k)}}_{\text{MS}}(\mc{P},D)$.
	\end{IEEEproof}
	
	\begin{claim}
There is at most one non-black element on the diagonal of $Q(1:(M-k)|1:(M-k))$.
\label{claim: 4}
\end{claim}
\begin{IEEEproof}
	Take an arbitrary $Q_{\hX|X} \in \mc{Q}^{{*}^{(k)}}_{\text{MS}}(\mc{P},D)$ satisfying all previous claims. Assume the contrary that there are at least two non-black diagonal elements $Q(i|i)$ and $Q(j|j)$, ($i \neq j, i,k \le M-k$). Thus, $Q(i|j)$ is red by Claim \ref{claim: 3}, which implies that there exists a $k \neq j$ where $Q(i|k)$ is black, because there should be a black element for each red element in a column. However, $k\neq i$ because we already know that $Q(i|i)$ is non-black. Thus, $Q(i|k)$ has to be an off-diagonal black element in $Q(1:(M-k)|1:M)$, which is contradictory to our first assumption of $Q$ satisfying all previous claims, including Claim \ref{claim: 1} and \ref{claim: 2}. Therefore, at most one diagonal element is non-black.
	\end{IEEEproof}

	\begin{claim}
	The only possible non-black element along the diagonal of $Q(1:(M-k)|1:(M-k))$ is the one corresponding to the smallest, or one of the smallest $D_i$s.
	\label{claim: 5}
	\end{claim}
		\begin{IEEEproof}
	Take an arbitrary $Q_{\hX|X} \in\mc{Q}^{{*}^{(k)}}_{\text{MS}}(\mc{P},D)$ satisfying all previous claims. Let the diagonal element in row $i, 1 \le i \le M-k$, be non-black. We show that for any $j \neq i, 1 \le j \le M-k$, $\hat{D}_j \ge \hat{D}_i$. By Claim \ref{claim: 3}, we know that any other off-diagonal entry in row $i$ of $Q(1:(M-k)|1:(M-k))$, including $Q(j|i)$ is red. We also know that $Q(j|j)$ is black and other entries in row $j$ of $Q(1:(M-k)|1:(M-k))$, including $Q(i|j)$, are either all red or all white due to Claims \ref{claim: 1} and \ref{claim: 2}. Thus, for all $1\le k\le M-k$ other than $i$ or $j$ we have $Q(k|j) \ge Q(k|i)$ because $Q(k|j)$ is either red or white, and $Q(k|i)$ is red, where both of them are in the same column. Since any row has to sum up to one, summing over rows $i$ and $j$ results in
	\begin{equation}
	Q(j|j) + Q(i|j) \le Q(i|i) + Q(j|i).
	\label{eq: claim 5 1}
	\end{equation}	
Besides, since $Q(j|j)$ is black, $Q(j|i)$ is either red or white, $Q(i|j)$ is red, and $Q(i|i)$ is either red or white, we have
\begin{equation}
\frac{Q(j|j)}{Q(j|i)} > \frac{Q(i|i)}{Q(i|j)}.
\label{eq: claim 5 2}
\end{equation}
We now show that 
	\begin{equation}
	1-\hat{D}_j = Q(j|j) \le Q(i|i) = 1-\hat{D}_i.
	\end{equation}
Assume the contrary that $Q(j|j) > Q(i|i)$. Thus, by \eqref{eq: claim 5 1} we have
\begin{equation}
0 < Q(j|j) - Q(i|i) \le Q(j|i) - Q(i|j),
\label{eq: i,j > j,i}
\end{equation}
which means $\frac{Q(j|j) - Q(i|i)}{Q(j|i) - Q(i|j)} \le 1$. However, \eqref{eq: claim 5 2} shows that
\begin{equation}
\frac{Q(j|j) - Q(i|i)}{Q(j|i) - Q(i|j)} \ge \frac{Q(j|j)}{Q(j|i)} > \frac{Q(i|i)}{Q(i|j)} \ge 0,
\end{equation}
which means
\begin{equation}
Q(j|j)-Q(i|i) > Q(j|i) - Q(i|j)
\label{eq: j,i > i,j}
\end{equation}
because $ e^ {\edp(Q)} = \frac{Q(j|j)}{Q(j|i)} > 1$, which contradicts \eqref{eq: i,j > j,i}.
\end{IEEEproof}

The claims above imply that for $D_1$ as one of the smallest $\{D_i\}_{i=1}^{M-k}$, all the other diagonal elements are black, and the non-zero elements in the row corresponding to $D_1$ are all red. This implies that for each red $Q(j |1), j \neq i$, there exists a diagonal element $1 - D_j$ in the same column $j$ which is $e^{\edp(Q)}$ times bigger than the red element $Q(j | 1)$. The proof is completed by summing over row $1$ entries and solving for $\edp$.
\end{IEEEproof}

Let $0\le k \le M-1$. Lemma \ref{lemma: epsilon_prime} provides a formula for the optimal $\edp(\cdot)$ among $(\mc{P},D)$-valid mechanisms with $k$ all-zero columns, in terms of their corresponding distortion values $\{D_i\}_{i=1}^{M-k}$. Besides, no mechanism with at least $k$ all-zero columns can be $(\mc{P},D)$-valid for $D < D^{(k)}$ due to Lemma \ref{lemma: not less than D^k}. Thus, for any $k$ and $D^{(k)} \le D \le D^{(k+1)}$, a lower bound on $\edpopt(\mc{P},D)$ can be derived by taking the minimum over $0 \le l \le k$ and all $(\mc{P},D)$-valid sets of $\{D_i\}_{i=1}^{M}$ that satisfy $(M-1-k)\frac{1-\frac{\sum_{i=2}^{M-k} D_i}{M-1-k}}{D_1} \ge 1$, or equivalently $\sum_{i=1}^{M-k} D_i \le M-1$. Moreover, for a mechanism with $k$ all-zero columns we have $D_i= 1$ for $i > M-k$, and thus $\{D_i\}_{i=1}^{M}$ can be $(\mc{P},D)$-valid if and only if $\{D_i\}_{i=1}^{M-k}$ is $(\mc{P},D-D^{(k)})$-valid, i.e. $\sum_{i=1}^{M-k} P_i D_i \le D- D^{(k)}$. This result in \eqref{eq: epsilon star k Class II} and completes the proof of the lower bound in Theorem \ref{theorem:2}.

We now proceed to the special case where $D < D^{(1)}$. For proving $\edpopt(\mc{P},D) \ge \log (M-1) \frac{1-D}{D}$, it suffices to show that $\epsilon^{*^{(0)}}_{\text{DP}}(\mc{P},D)$ is greater than or equal to $\log(M-1)\frac{1-D}{D}$. We need the following Lemma.

\begin{lemma}
	Let $\{a_i\}_{i=1}^n$, $\{b_i\}_{i=1}^n$, and $\{a'_i\}_{i=1}^n$ be a collection of real numbers between $0$ and $1$, such that
	\begin{equation}
	\sum_{i=1}^{n} a_i = \sum_{i=1}^{n} a'_i,
	\end{equation}
	and $b_1 \le b_2 \le \ldots \le b_n$.
	If $a'_1 \ge a_1$ and $a'_i \le a_i$ for $i=2,3,\cdots,n$, then
	\begin{equation}
	\sum_{i=1}^{n} a'_i b_i \le \sum_{i=1}^{n} a_i b_i.
	\end{equation}
	\label{lemma: a' vs a}
\end{lemma}

Then, assume the contrary that for some $D < D^{(1)}$, all mechanisms in $\Qopt (\mc{P},D)$ achieve a strictly smaller $\edp(\cdot)$ than $\log (M-1)\frac{1-D}{D}$. From Lemma \ref{lemma: epsilon_prime} we know that there exists an optimal mechanism $Q_{\hX|X}$ with the set of distortions $\left\{D_i\right\}_{i=1}^{M}$, such that $\edp(Q)$ is given by \eqref{eq: epsilon_prime}, and without loss of generality $D_1 \le D_2 \le \cdots \le D_M$ due to Lemma \ref{lemma: reverse order of distortion and probability}. Hence, the contrary assumption is that there exists an optimal $Q_{\hX|X}$ such that
\begin{subequations}
	\begingroup
	\allowdisplaybreaks
	\begin{align}
	\edpopt(\mc{Q},D)=\edp(Q_{\hX|X}) &= \log (M-1)\frac{1-\frac{\sum_{i=2}^{M} D_{i}}{M-1}}{D_{1}} \\
	&< \log (M-1) \frac{1-D}{D}.
	\end{align}
	\endgroup
\end{subequations}
Thus:
\begin{equation}
(1-D) D_{1} + \frac{D}{M-1} \sum_{i=2}^{M}  D_{i} > D.
\label{eq: contradiction 1}
\end{equation}
On the other hand, since $Q_{\hX|X}$ is supposed to satisfy the distortion constraint for any $P \in \mc{P}$, including ${P^{*}} = {\argmax}_{P \in \mc{P}} P_M$, we have
\begin{subequations}
	\begingroup
	\allowdisplaybreaks
	\begin{align}
	\sum_{i=1}^{M} P^{*}_i D_i &= P^{*}_1 D_{1} + \sum_{i=2}^{M} P^{*}_{i} D_{i}\\
	\label{eq: contradiction 2}
	&= (1-D) D_{1} + \frac{D}{M-1} \sum_{i=2}^{M} D_{i} \le D,
	\end{align}
	\endgroup
\end{subequations}
where \eqref{eq: contradiction 2} is in view of Lemma \ref{lemma: a' vs a} and the fact that $D \le D^{(1)}$ implies $P^{*}_i \ge \frac{D}{M-1}$ for $i=2,3,\cdots,M$ and $P^{*}_1 \le 1-D$. Obviously \eqref{eq: contradiction 2} contradicts \eqref{eq: contradiction 1}. Thus, for $0 \le D \le D^{(1)}$, we have
\begin{equation}
\edpopt(\mc{Q},D) \ge \log (M-1) \frac{1-D}{D}.
\label{eq: LB for d less than d_1}
\end{equation}

\subsubsection{Achievability for Theorem \ref{theorem:2}}

First, we show that for $0 \le D < D^{(1)}$, the optimal leakage in \eqref{eq: LB II} is achievable. Consider the following mechanism 
\begin{equation}
Q(j|i) = \begin{cases}
1-D, & i=j,\\
\frac{D}{M-1}, & i \neq j.
\end{cases}
\end{equation}
Observe that $Q$ is $(\mc{P},D)$-valid and $\edp(Q)=\log (M-1) \frac{1-D}{D}$. Therefore, the lower bound in \eqref{eq: LB for d less than d_1} is tight and $\edpopt(\mc{P},D)=\log (M-1) \frac{1-D}{D}$ for $0 \le D < D^{(1)}$.

We now prove that the lower bound in \eqref{eq: LB II} is achievable for $D^{(1)} \le D < D^{(M-1)}$.
To this end, we construct the following mechanism. For any given $0 \le k \le M-1$ and the optimal set $\{D^*_i\}_{i=1}^{M-1}$ in \eqref{eq: epsilon star k Class II}, consider the following mechanism.
\begin{equation}
Q^{(k)^{*}}(j|i) = \begin{cases}
1-D^*_i, & i=j \le M-k,\\
D_i \frac{1-D^*_j}{\sum_{l\neq i} 1-D^*_l}, & i \neq j, \;\; i,j < M-k,\\
Q^{(k)^{*}}(j|M-k), & i > M-k, j \le M-k,\\
0, & j>M-k.
\end{cases}
\end{equation}
We now verify that $\edp(Q^{(k)^{*}}) = (M-1-k) \frac{1-\frac{\sum_{i=2}^{M-k} D^*_i}{M-1-k}}{D^*_1}$.
Since each of the last $k$ rows in the above matrix are equal to the ${(M-k)}^{th}$ row, and the last $k$ columns are all equal to zero, it suffices to check the $\edp(\cdot)$ for the square matrix formed by the first $M-k$ rows and columns. The ratio of any two elements in the same column in $Q^{(k)^{*}}$ belongs to the set $\{c_1,c_2, \ldots, c_{M-k}\}$, where 
\begin{equation}
c_i=(M-1-k)\frac{1-\frac{\sum_{j\neq i} D_j}{M-1-k}}{D_i}.
\end{equation}
From Lemma \ref{lemma: reverse order of distortion and probability} we have that $D_1 \le \ldots \le D_{M-k}$, which in turn implies that for any $0 \le j \le M-k$:
\begin{equation}
(M-1-k)\frac{1-\frac{\sum_{i=2}^{M-k} D_i}{M-1-k}}{D_1} \ge (M-1-k)\frac{1-\frac{\sum_{i\neq j} D_i}{M-1-k}}{D_j}.
\end{equation}
Therefore, $\edp(Q^{(k)^{*}}) = (M-1-k)\frac{1-\frac{\sum_{i=2}^{M-k} D_i}{M-1-k}}{D_1}$.

\subsection{Proof of Theorem \ref{theorem:3}: Class III source sets\label{proof:theorem:3}}

\begin{IEEEproof}
Recall that a Class III source set $\mc{P}$ can be written as a union of Class II source sets as $\mc{P} =  \underset{T \in\mc{T}_{\mc{P}}}{\cup} \mc{P}|_{T}$. Furthermore, each of these partitions $\mc{P}|_{T}$ can be mapped to $\mc{S}_0$ with the appropriate permutation to get $\overline{\mc{P}|_{T}}$. We write the intersection and union of the mapped partitions as $\mc{P}^{\cap}$ and $\mc{P}^{\cup}$, respectively. Since $\mc{P}^{\cap}$ and $\mc{P}^{\cup}$ are Class II source sets, we can compute the optimal leakage and the corresponding mechanism for these two sets. Moreover, mapping $\mc{P}^{\cup}$ and $\mc{P}^{\cap}$ back into the original partitions results in two sets $\mc{P}^{\text{UB}}$ and $\mc{P}^{\text{LB}}$ that contain and are contained in $\mc{P}$, respectively.

Formally, let 
\begin{subequations}
	\begingroup
	\allowdisplaybreaks
\begin{align}
\mc{P}^{\text{LB}} &= \cup_{T \in \mc{T}_{\mc{P}}} T^{-1}(\mc{P}^{\cap}),\\
\mc{P}^{\text{UB}} &= \cup_{T \in \mc{T}_{\mc{P}}} T^{-1}(\mc{P}^{\cup}).
\end{align}
	\endgroup
\end{subequations}

For the $\mc{P}$ shown in Figure \ref{fig: folded 1} and the corresponding $\mc{P}^{\cup}$ and $\mc{P}^{\cap}$ in Figure \ref{fig: folded 2}, Figure \ref{fig: folded 3} below illustrates the $\mc{P}^{\text{LB}}$ and $\mc{P}^{\text{UB}}$.
	\begin{figure}[H]
		\centering
		\includegraphics[width=0.6\columnwidth]{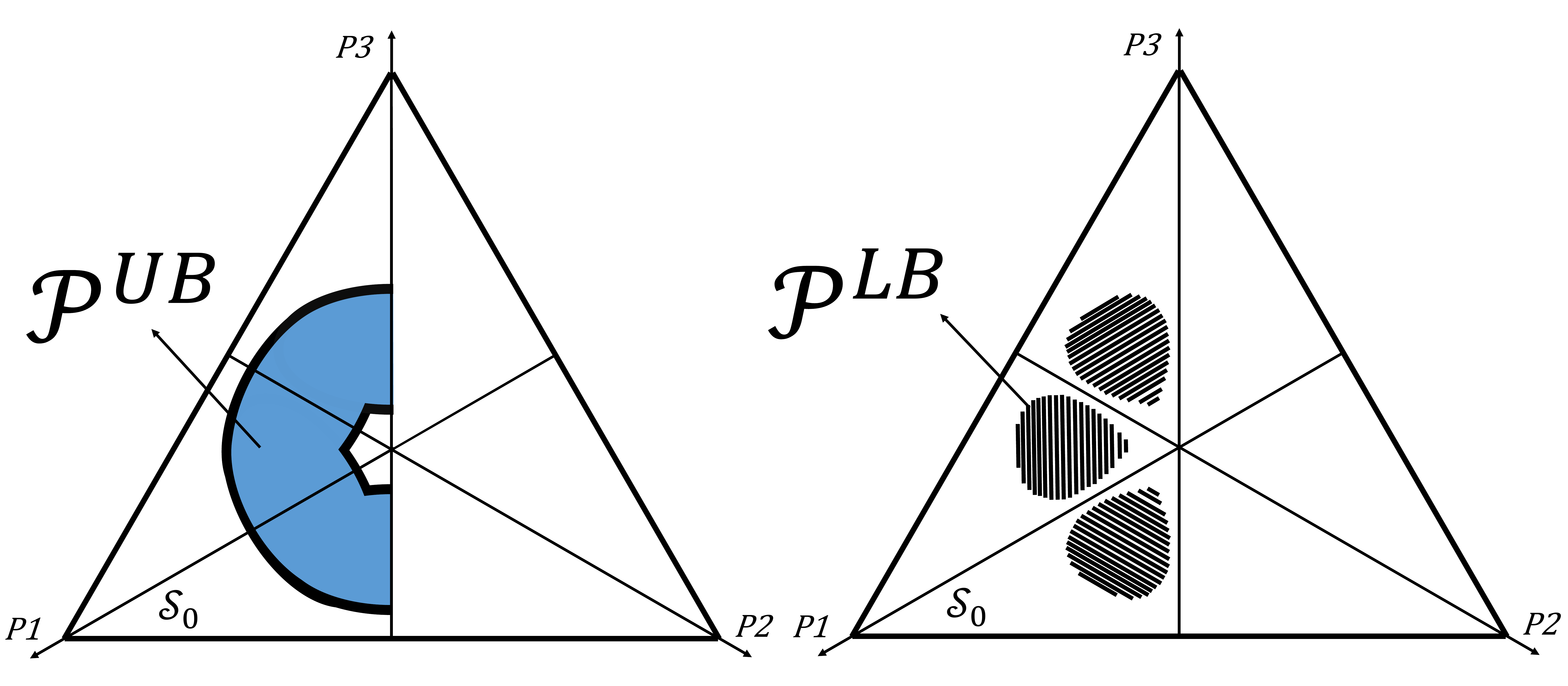}
		\caption{A source set $\mc{P}$ and its folded versions.}
		\label{fig: folded 3}
	\end{figure}
Recall that for two sets $\mc{P}_1$ and $\mc{P}_2$, if $\mc{P}_1 \subseteq \mc{P}_2$, then $\edpopt(\mc{P}_1,D) \ge \edpopt(\mc{P}_2,D)$. Thus, it suffices to show the following:
\begin{enumerate}[label=(\roman*)]
	\item $\mc{P}^{\text{LB}} \subseteq \mc{P} \subseteq \mc{P}^{\text{UB}}$, and
	\item The optimal DP leakage for $\mc{P}^{\text{LB}}$ and $\mc{P}^{\text{UB}}$ are given by
	\begin{align}
	\edpopt (\mc{P}^{\text{LB}},D) &= \epsilon^*_{\text{DP:III}} (\mc{P}^{\cap},D,\mc{T}_\mc{P}),\label{eq: proof for LB and cap}\\
	\edpopt (\mc{P}^{\text{UB}},D) &= \epsilon^*_{\text{DP:III}} (\mc{P}^{\cup},D,\mc{T}_\mc{P})\label{eq: proof for UB and cup},
	\end{align}
	where $\epsilon^{*}_{\text{DP:III}}(\cdot)$ is defined in \eqref{eq: LB III}.
\end{enumerate}

\textit{Proof of (i):} For each $T \in \mc{T}_\mc{P}$, since $\mc{P}^{\cap} \subseteq \overline{\mc{P}|_{T}}$ we have $T^{-1} (\mc{P}^{\cap}) \subseteq \mc{P}|_{T}$. After taking union over all $T \in \mc{T}_\mc{P}$, we have $\mc{P}^{\text{LB}} \subseteq \mc{P}$. One can immediately show that $\mc{P} \subseteq \mc{P}^{\text{UB}}$.

\textit{Proof of (ii):}
We first prove \eqref{eq: proof for LB and cap}. A similar argument proves \eqref{eq: proof for UB and cup}. Recall that $\mc{P}^{\text{LB}} = \cup_{T \in \mc{T}_{\mc{P}}} T^{-1}(\mc{P}^{\cap})$, and thus, for any $P \in \mc{P}^{\cap}$ and $T \in \mc{T}_{\mc{P}}$ we have $T^{-1}(P) \in \mc{P}^{\text{LB}}$.
This means that for any given $(\mc{P}^{\text{LB}},D)$-valid mechanism $Q_{\hX|X}$, $Q_{T(\hX)|T(X)}$ is also $(\mc{P}^{\text{LB}},D)$-valid and $\edp(Q_{\hX|X}) = \edp( Q_{T(\hX)|T(X)})$. Since $\edp(Q)$ is a quasi-convex function of $Q$ due to Lemma \ref{lemma: convexity of epsilon}, there exists an optimal mechanism achieving $\edpopt(\mc{P}^{\text{LB}},D)$ for which
	\begin{equation}
	D_{T(i)}= D_i, \; \text{ for any } T \in \mc{T}_{\mc{P}}, i=1,\ldots,M.
	\label{eq: extra condition for III}
	\end{equation}
Hence, it suffices to search over only those $(\mc{P}^{\cap},D)$-valid mechanism that satisfy \eqref{eq: extra condition for III}  in order to find $\epsilon^{(k)^{*}}_{\text{DP}}(\mc{P}^{\text{LB}},D)$. Furthermore, since $\mc{P}^{\cap}$ is a Class II source set, we can use the results from Theorem \ref{theorem:2}. We now show $\edpopt (\mc{P}^{\text{LB}},D) = \epsilon^*_{\text{DP:III}} (\mc{P}^{\cap},D,\mc{T}_\mc{P})$.

First consider the case where $D \ge \frac{M-1}{M}$. Clearly, choosing $Q(i|j) = \frac{1}{M}$ achieves $\edpopt(\mc{P}^{\text{LB}}, D) = 0$, while the distortion constraint is also satisfied.

For $D < \frac{M-1}{M}$, similar to the proof for Theorem \ref{theorem:2}, we first restrict the set of mechanisms to those that have a fixed number $k$ of all-zero columns, $k=0,1,\ldots,M-1$.
For any such $k$, the optimal leakage is given by \eqref{eq: epsilon star k Class III}, where the third constraint is a result of \eqref{eq: extra condition for III}.
Note that \eqref{eq: epsilon star k Class III} results from the addition of the constraint in \eqref{eq: extra condition for III} to the constraints in \eqref{eq: epsilon star k Class II} for a Class II source set.
The optimal $\edpopt(\mc{P}^{\text{LB}}, D)$ is then the minimum of $\epsilon^{(k)^*}_{\text{DP:III}} (\mc{P}^{\cap},D,\mc{T}_\mc{P})$ over all $k$, resulting in \eqref{eq: LB III}.

Finally, for $D < D^{(1)}$, analogous to Theorem \ref{theorem:2}, we can still show that the optimal mechanism is symmetric. Recall that for a Class II source set $\mc{P}^{\cap}$ and $D < D^{(1)}$, the optimal mechanism achieving $\edpopt(\mc{P}^{\cap},D)$ is symmetric, and thus, does not violate \eqref{eq: extra condition for III}. Hence, we have 
\begin{equation*}
\epsilon^*_{\text{DP:III}} (\mc{P}^{\cap},D,\mc{T}_\mc{P})  = \edpopt(\mc{P}^{\cap},D) = \log \left((M-1) \frac{1-D}{D}\right).
\end{equation*}
\end{IEEEproof}

Note that in contrast to Theorem \ref{theorem:2}, we no longer have distortion thresholds $D^{(2)}, D^{(3)}, \ldots, D^{(M-2)}$, where in each of them only mechanisms with specific number of all-zero columns are allowed. This is due to the constraint in \eqref{eq: extra condition for III}, which may not allow a gradual shrinking of output support set.

\section{Conclusion}
In this paper, we have quantified the privacy-utility tradeoffs for a dataset under different assumptions on distribution knowledge (classes) and for Hamming distortion using differential privacy as the leakage metric. The guarantees we can make under differential privacy are stronger than those under mutual information-based measures of privacy leakage: DP leakage is lower bounded by MI leakage.
We divide source sets into three classes. For Class I the optimal mechanism is symmetric. For Class II achieving optimal leakage involves reducing the output space as the distortion increases. For Class III sets we can use Class II results to develop upper and lower bounds on the leakage.


Our results show that symmetric distortion, such as randomized response~\cite{Warner}, is optimal when very little is known about the source distribution or when the distortion requirement is very strict. In cases where the source distribution is partially known, data publishers can take advantage of this to tailor a local privacy mechanism to guarantee lower privacy leakage for the same distortion, or lower distortion for the same privacy leakage. These gains can be significant if quite a lot is known about the source, such as Class II sources, and degrades as less and less information is known. These results imply that domain knowledge or public data should be used when designing mechanisms for publishing private data.


There are several interesting questions which we leave for future work. Obviously, a full characterization of Class III sources would be welcome, but the techniques here should extend directly to general discrete distortion measures (linear and nonlinear). Extensions to continuous source distributions may be trickier, but perhaps a starting point would be distributions with bounded support. Finally, understanding the implications of this simple model to categorical and hierarchically categorized data would help build insight into designing practical source-aware data release mechanisms.


\bibliographystyle{IEEEtran}

\bibliography{Privacy}

\end{document}